\documentclass[11pt]{article} 

\usepackage{jheppub}

\usepackage{amsmath,amsfonts,amsbsy,amssymb,array,accents,dsfont}
\usepackage{enumerate}
\usepackage[shortlabels]{enumitem}
\usepackage{graphicx} 
\usepackage{upgreek}
\usepackage{empheq}
\usepackage[many]{tcolorbox}
\usepackage{bm}
\usepackage{cancel}
\usepackage{slashed}
\usepackage{nccmath}
\usepackage{tikz}

\newtcolorbox{empheqboxed}{colback=gray!30, 
 colframe=white,
 width=\textwidth,
 sharpish corners,
 top=-2mm, 
 bottom=0pt
}

\let\includefigures=\iftrue
\let\useblackboard==\iftrue
\newfam\black

\PassOptionsToPackage{ 
     colorlinks=true,
     linkcolor=myBlue,
     urlcolor=blue,
     filecolor=black,
     citecolor=myRed,
}{hyperref}

\usepackage{xparse}
\usepackage{xspace}

\NewDocumentCommand\eqn{om}{%
  \IfNoValueTF{#1}
     {\[ #2 \]}
     {\begin{equation}\label{#1} #2  \end{equation} \expandafter\newcommand\csname #1\endcsname{\eqref{#1}\xspace}\ignorespaces}
}
\NewDocumentCommand\eqna{om}{%
  \IfNoValueTF{#1}
    {\begin{align*} #2 \end{align*}}
    {\begin{equation}\label{#1}\begin{split} #2  \end{split}\end{equation} \expandafter\def\csname #1\endcsname{\eqref{#1}\xspace}\ignorespaces}
}

\newcommand{\rcite}{\cite}

\def\pup{{p}}
\def\thetaup{{\theta}}
\def\gammaup{{\lambda}}
\def\betaup{{\pi}}

\def\bstr{{\mathbi{b}_s}}
\def\cstr{{\mathbi{c}_s}}
\def\bnul{{\mathbi{b}_n}}
\def\cnul{{\mathbi{c}_n}}

\def\cstrb{{\bar{\mathbi{c}}_s}}

\def\cnulb{{\bar{\mathbi{c}}_n}}

\def\betattil{{{\hat\betab_n}}}
\def\gammattil{{{\hat\gammab_n}}}

\def\gammattilb{{{\hat{\bar\gammab}_n}}}
\def\betanul{\betab_n}
\def\gammanul{\gammab_n}
\def\betastr{\betab_s}
\def\gammanulb{\bar{\gammab}_n}

\def\gammastr{\gammab_s}

\def\phinul{{\phib_n}}
\def\chinul{{\chib_n}}
\def\phistr{{\phib_s}}
\def\phistrbar{{\bar\phib_s}}
\def\chistr{{\chib_s}}

\def\xistr{{\xib_s}}

\def\xinul{{\xib_n}}

\def\betab{{\boldsymbol\beta}}
\def\gammab{{\pmb\gamma}}

\def\rhob{{\boldsymbol\rho}} 
\def\sigmab{{\boldsymbol\sigma}} 
 
\def\phib{{\boldsymbol\phi}} 
\def\xib{{\boldsymbol\xi}} 
 
\def\phib{{\boldsymbol\phi}} 
\def\chib{{\boldsymbol\chi}}

\def\sl{\text{sl}}
\def\su{\text{su}}

\def\vareps{\varepsilon}

\def\str{{\rm str}}

\def\bos{{\rm bos}}
\def\fer{{\rm fer}}
\def\flat{{\rm flat}}

\def\sltwo{{\mathfrak{sl}(2,\mathds R)}}
\def\SLtwo{{\ensuremath{SL(2,\mathds R)}}}

\def\sutwo{{\mathfrak{su}(2)}}
\def\SUtwo{{SU(2)}}
\def\PSU{{PSU(1,1|2)}}
\def\psu{{\mathfrak{psu}(1,1|2)}}

\def\Uone{U(1)}
\def\uone{{\mathfrak{u}(1)}}

\def\mathbi#1{\textbf{\em #1}}
\def\tight#1{\! #1 \!}  

\def\({\left(}
\def\){\right)}
\def\[{\left[}
\def\]{\right]}

\def\ie{{i.e.}}
\def\eg{{e.g.}}

\def\etc{{etc}}

\def\brst{{\sf brst}}
\def\tot{{\rm tot}}
\def\con{{\rm con}}
\def\nul{{\rm null}}
\def\mix{{\rm mixed}}

\def\min{{\rm min}}

\def\gstr{g_{\textit s}^{\;}}

\def\nfive{{n_5}}

\DeclareMathSymbol{\medhatsym}{\mathord}{largesymbols}{"62} 

\DeclareMathSymbol{\medtildesym}{\mathord}{largesymbols}{"65}

\makeatletter
\newcommand*\rel@kern[1]{\kern#1\dimexpr\macc@kerna}
\newcommand*\widebar[1]{%
  \begingroup
  \def\mathaccent##1##2{%
    \rel@kern{0.8}%
    \overline{\rel@kern{-0.8}\macc@nucleus\rel@kern{0.2}}%
    \rel@kern{-0.2}%
  }%
  \macc@depth\@ne
  \let\math@bgroup\@empty \let\math@egroup\macc@set@skewchar
  \mathsurround\z@ \frozen@everymath{\mathgroup\macc@group\relax}%
  \macc@set@skewchar\relax
  \let\mathaccentV\macc@nested@a
  \macc@nested@a\relax111{#1}%
  \endgroup
}
\makeatother

\def\half{\frac12}
\def\coeff#1#2{{\textstyle \frac{#1}{#2}}}
\def\hf{\coeff12}
\def\tr{{\rm Tr}}

\def\One{{\hbox{1\kern-1mm l}}}

\def\barray{\begin{array}}
\def\earray{\end{array}}
\def\be{\begin{equation}}
\def\ee{\end{equation}}
\def\bea{\begin{eqnarray}}
\def\eea{\end{eqnarray}}
\def\bal{\begin{align}}
\def\eal{\end{align}}
\def\nn{\nonumber}

\newcommand{\bN}{{\mathbb N}}

\newcommand{\bR}{{\mathbb R}}
\newcommand{\bS}{{\mathbb S}}
\newcommand{\bT}{{\mathbb T}}

\newcommand{\bZ}{{\mathbb Z}}

						\def\sfH{{\mathsf H}}
		\def\sfJ{{\mathsf J}}

\def\sfi{{\mathsf i}}		\def\sfj{{\mathsf j}}		\def\sfk{{\mathsf k}}		
\def\sfm{{\mathsf m}}		\def\sfn{{\mathsf n}}				
						
		\def\sfv{{\mathsf v}}

		\def\mfJ{{\mathfrak J}}

\def\mfa{{\mathfrak a}}

\definecolor{cardinal}{rgb}{0.6,0,0}
\definecolor{darkgreen}{rgb}{0,0.4,0}
\definecolor{green}{rgb}{0,0.4,0}
\definecolor{golden}{rgb}{0.92, 0.7, 0}
\definecolor{midnight}{rgb}{0, 0, 0.5}
\definecolor{darkblue}{rgb}{0, 0, 0.7}

\numberwithin{equation}{section}

\mathchardef\mhyphen="2D

  \def\cC{\mathcal {C}}
  
\def\cG{\mathcal {G}} \def\cH{\mathcal {H}} 
\def\cJ{\mathcal {J}}  
\def\cM{\mathcal {M}} \def\cN{\mathcal {N}} \def\cO{\mathcal {O}}
  \def\cR{\mathcal {R}}
\def\cS{\mathcal {S}} \def\cT{\mathcal {T}}

\def\one{{\hbox{\kern+.5mm 1\kern-.8mm l}}}
\def\zero{{\hbox{0\kern-1.5mm 0}}}

\newcommand{\ket}[1]{{\,| {#1} \rangle}}

\def\id{\textrm{id}}

\def\id{{1 \kern-.28em {\rm l}}}

\def\journal#1&#2(#3){\unskip, \sl #1\ \bf #2 \rm(19#3) }
\def\andjournal#1&#2(#3){\sl #1~\bf #2 \rm (19#3) }

\def\ie{{\it i.e.}}
\def\eg{{\it e.g.}}

\def\etc{{\it etc}}

\def\coeff#1#2{{\textstyle{\frac{#1}{ #2}}}}
\def\half{\frac12}
\def\hf{{\textstyle\half}}
\def\ket#1{|#1\rangle}

\def\One{{1\hskip -3pt {\rm l}}}

\catcode`\@=11
\def\slash#1{\mathord{\mathpalette\c@ncel{#1}}}
\def\vareps{\varepsilon}
\def\underrel#1\over#2{\mathrel{\mathop{\kern\z@#1}\limits_{#2}}}

\catcode`\@=12

\def\ket#1{\left| #1\right\rangle}

\def\exp{{\rm exp}}
\def\gh{{\rm gh}}
\def\m{{\rm m}}

\def\ie{{\it i.e.}}
\def\eg{{\it e.g.}}

\title{
On the string theory of a single NS5-brane}

\author{
Andrea Dei {\it and} Emil J. Martinec
}

\affiliation{
\vskip 0.01cm
Kadanoff Center for Theoretical Physics, Enrico Fermi Institute, and Department of Physics\\ 
University of Chicago,
5640 S. Ellis Ave.,
Chicago IL 60637\\ 
}

\emailAdd{%
adei@uchicago.edu,
e-martinec@uchicago.edu}

\abstract{We formulate a worldsheet description of string theory in the background of a single decoupled NS5-brane, which is a particularly simple example of ``little string'' holography.
The worldsheet theory involves a gauged Wess-Zumino-Witten model for the group coset $\cG/\cH$, where $\cG = \PSU\times \bR^{5,1}$ and $\cH=\Uone\times\Uone$.  
A variant of the embedding of $\cH$ into $\cG$ describes a background that interpolates between the fivebrane throat in the UV and $AdS_3$ in the IR, which is dual to the (single trace) $T\bar T$-deformed symmetric product orbifold $(\bT^4)^N/S_N$. 
Well-known difficulties with a worldsheet theory for $n_5=1$ fivebranes are bypassed by the incorporation of an extra constraint on the dynamics.  This constraint arises naturally from consideration of the effects of deforming the theory by one of the four Ramond-Ramond moduli of the spacetime background, which for $n_5=1$ imposes the constraint that our analysis has identified. 
Along the way, we construct a map between the supergroup formulation of the $n_5=1$ worldsheet theory and the more conventional RNS worldsheet formalism.
}

\begin{document}
\maketitle
\hypersetup{pageanchor=true}
\pagenumbering{arabic}


\section{Introduction and summary of results} 
\label{sec:intro}

Neveu-Schwarz fivebranes (NS5) are solitonic objects in string theory, the magnetic duals of fundamental strings.  As such, the string interaction strength grows to infinity as strings approach fivebranes, unless some dynamical phenomenon prevents them from doing so.  The near-source limit of a collection of coincident $\nfive$ NS5-branes has a long throat, along which the string dilaton which characterizes the local string interaction strength naively grows without bound.

The decoupling limit of fivebranes, in which one scales to zero the asymptotic value of the string coupling $g_s$ while scaling the fivebranes' separation to zero commensurately, is of special interest because there is a holographic duality between gravitational dynamics in the fivebrane throat and a non-gravitational dual theory which is {\it not} a quantum field theory.  Instead, the dual is a non-abelian, strongly coupled non-gravitational string theory known as {\it little string theory} (LST).  The little string's tension is parametrically lighter than that of the fundamental string, 
\be
\label{alplittle}
\alpha'_{\it little} = \nfive\alpha' ~.
\ee
The Hagedorn thermodynamics of the little string, having the equation of state
\be
S=2\pi\sqrt{\alpha'_{\it little}}\, E ~,
\ee
maps to the thermodynamics of black fivebranes at high energy.%
\footnote{The qualitative picture of little strings is duality frame dependent.  In type IIA, one can think of them as D2-branes stretching between NS5-branes; in type IIB, instantons in the low-energy 5+1d Yang-Mills gauge theory are codimension four objects which are effective strings.  Either way, they are inherently related to non-abelian NS5-brane dynamics. }  

This setup is particularly interesting because it takes us beyond the usual AdS/CFT paradigm, as the throat geometry is not asymptotically Anti-de Sitter and has some features similar to those of flat spacetime (see for instance~\rcite{Marolf:2006bk,Marolf:2007ys, Giveon:2017myj, Asrat:2017tzd, Chang:2023kkq}). However, because of the presence of an object lighter than the fundamental string, the dynamics is inherently strongly coupled.   
It is thus hard to make headway.

One way of bypassing the non-abelian strong coupling dynamics of little strings is to slightly separate the fivebranes, which gives mass to the little string excitations~\rcite{Giveon:1999px,Giveon:1999tq}.
Fundamental strings are then prevented from approaching the fivebrane sources arbitrarily closely; the string coupling saturates at a scale governed by the fivebrane separation, leading to a self-consistent perturbative expansion (for a recent discussion, see~\rcite{Martinec:2024emf}).
String perturbation theory is then valid below this mass scale.\footnote{For computations of IR observables, see for example \rcite{Aharony:2004xn,Chang:2014jta,Lin:2015zea}.}  

A particularly symmetric arrangement of fivebranes leads to a self-consistently weakly coupled and solvable worldsheet description built on gauged Wess-Zumino-Witten (WZW) models~\rcite{Giveon:1999px,Giveon:1999tq}.  A variant, equivalent formulation of the gauging is
\be
\label{GmodH}
\frac\cG\cH = \frac{\SLtwo\times\SUtwo}{\Uone\times\Uone}  ~,
\ee
where the current algebra level of each factor of $\cG$ is $\nfive$, and $\cH=\Uone\times\Uone$ is a pair of isometries of $\cG$ generated by null Killing vectors~\rcite{Israel:2004ir,Martinec:2017ztd}. While this setup bypasses the issue of strongly coupled dynamics, it only describes the low-energy excitations in a perturbative expansion around a particular Coulomb branch configuration, and is not a fundamental definition of the theory.

\bigskip

We can overcome the obstacles of non-abelian and strong coupling dynamics by considering a single fivebrane.  For $\nfive=1$, there is no fractionation of the string tension~-- the little string inside the fivebrane {\it is} the fundamental string. It is no longer true that there is an object lighter than the fundamental string, and there is no excuse not to be able to solve the theory. Moreover, abelianizing the dynamics, one may hope that the boundary theory is simple enough to identify a first principles definition of little string theory, at least in this regime. 

However, one encounters a further technical challenge~-- the seeming absence of a unitary supersymmetric $\SUtwo$ WZW model at level $\nfive=1$.  In the RNS formulation of worldsheet string theory, unitarity seems to place a lower bound of $\nfive=2$, since even if the underlying bosonic $SU(2)$ sigma model is trivial (level zero), the fermions needed for worldsheet supersymmetry give a contribution of two to the current algebra level.

This issue arises already in a closely related situation, namely string theory on $AdS_3\times \bS^3\times\bT^4$ sourced by a single fivebrane and a large collection of fundamental strings, which employs the same sigma model on $\cG$.  There, a solution to this problem was found~\rcite{Eberhardt:2018ouy} 
through an alternative ``hybrid'' worldsheet formalism, that uses the superalgebra $\psu$ at level $\nfive=1$ in the target space description rather than the standard Ramond-Neveu-Schwarz (RNS) formulation, which relies on a worldsheet supersymmetric sigma model on $\cG$.

The existence of this hybrid formulation for $\nfive>1$ dates back earlier, to the work of~\rcite{Berkovits:1999im}.  There, both RNS and hybrid formulations exist, and are related by bosonization of the various worldsheet fermions (as well as spinor ghosts), followed by a field redefinition.  We refer to this transformation as the RNS\,$\leftrightarrow$\,hybrid map.%
\footnote{This map is similar in spirit to the relation between the RNS and Greeen-Schwarz formulations of flat spacetime string theory in light-cone gauge, see for instance~\rcite{Green:1987sp,Polchinski:1998rr} for reviews.} 
One of our main results is to provide such a map for $\nfive=1$.%
\footnote{Recently,~\rcite{Eberhardt:2025sbi} suggested that an RNS formulation might exist, but gave no concrete mechanism for overcoming the apparent lack of unitarity.} 

\bigskip

Our analysis brings together all of these threads: The use of the hybrid formalism to describe $\nfive=1$; the use of null gauging to provide a worldsheet description of the fivebrane throat; and the use of the RNS\,$\leftrightarrow$\,hybrid map to export the null gauging construction to $\nfive=1$. 
The analysis proceeds in three steps:

\paragraph{Step one:} After reviewing in Section~\ref{sec:hybrid kgt1} the standard $n_5>1$ hybrid formalism for $AdS_3 \times S^3 \times \mathbb T^4$ string theory, we provide in Section~\ref{sec:levelone} a new hybrid formulation at $n_5=1$, which we show is equivalent to the existing version given in~\rcite{Eberhardt:2018ouy,Gaberdiel:2022als,Dei:2023ivl}.  
The basic idea is to enlarge the field content and at the same time introduce an additional constraint that projects down to the minimal field content.  The enlarged field content admits an RNS $\leftrightarrow$ hybrid~map, whereas the original formulations of~\rcite{Eberhardt:2018ouy,Gaberdiel:2022als,Dei:2023ivl} do not.

\paragraph{Step two:} In Section~\ref{sec:nullhybrid}, we use this RNS\,$\leftrightarrow$\,hybrid map to translate into hybrid variables the RNS formulation of little string theory at $\nfive\ge2$ as a null-gauged WZW model. 

\paragraph{Step three:} In Section~\ref{sec:nullgauge keq1}, we assemble the appropriate BRST operator, including the additional constraint, that extends the construction to $n_5=1$ and define the string theory of a single, decoupled NS5-brane. 

\vskip .5cm

The constraint introduced in the first step seems rather peculiar at first sight.  For instance, in the RNS formulation it imposes a relation between the spin fields that realize fermionic currents, which are used to build the spacetime $\cN=(4,4)$ superconformal symmetry of the background.  As such, one might think that this construction should break some of the spacetime supersymmetry; and yet, the hybrid formulation employed in the existing literature for $\nfive=1$~\rcite{Eberhardt:2018ouy, Gaberdiel:2021njm, Gaberdiel:2022als, Dei:2023ivl} has the full complement of $\cN=(4,4)$ superconformal symmetries in spacetime.

There are yet more peculiarities of our formulation.
We employ a free field construction for the worldsheet dynamics throughout most of our analysis.  However, at step one above, the enlarged field content includes a bosonic $\sltwo$ current algebra at level three.  The corresponding WZW model is only asymptotically free near the $AdS_3$ boundary.  The WZW model is not free, so how can we get away with a free field BRST formulation of the extra constraint? 

A resolution of these puzzles lies in a consideration of the effect of turning on the Ramond-Ramond (R-R) moduli of the background.  In the context of $\nfive>1$, the effect of these moduli is quite dramatic~\rcite{Seiberg:1999xz}.  
The $\SLtwo$ WZW model which describes string theory on $AdS_3$ has a continuum of ``long string'' states~\rcite{Maldacena:2000hw}.  Turning on R-R moduli generically erects an asymptotic ``wall'' that confines strings to the $AdS_3$ interior, and thus lifts the continuum~\rcite{Seiberg:1999xz}.  

The $\PSU$ WZW sigma model as formulated in~\rcite{Gotz:2006qp} naturally has an R-R  deformation turned on, and so describes the theory at a somewhat different point in the moduli space than one finds in the RNS formulation.%
\footnote{One can of course work at the same point as the RNS formalism for $\nfive\ge 2$, with no R-R moduli turned on, by simply following the $\SLtwo\times\SUtwo$ WZW theory through the RNS $\rightarrow$ hybrid map.  For $\nfive=1$ this seems not to be possible, as one would then encounter the pathologies that heretofore have plagued the RNS formulation at $\nfive=1$.}

For $\nfive=1$, all the physical states are concentrated near the $AdS_3$ boundary, and the effect of this deformation is dramatically different~-- it decouples a linear combination of fermions, having the same effect as the constraint that imposes this decoupling in the free field approximation.  The fact that the deformation is an $\cN=(4,4)$-preserving modulus explains why a constraint that eliminates from the worldsheet theory some of the free fields which realize spacetime supersymmetry, nevertheless manages to preserve the full complement of spacetime supersymmetries, in terms of composite operators in the effective dynamics.  The analysis explains how the use of free field theory fits within the larger context of dynamics on $AdS_3$, as an asymptotic approximation to the full dynamics.  And since this free field dynamics matches that of the dual spacetime CFT~\rcite{Eberhardt:2018ouy,Eberhardt:2019ywk}, we can be confident that it is correct. 

\vskip .7cm
The uses of null gauged WZW models extend well beyond the description of NS5-branes separated on their Coulomb branch~\rcite{Martinec:2017ztd}.  One can alter the embedding of the gauge group $\cH$ so that it lies partly in the $\SLtwo\times\SUtwo$ describing directions transverse to the fivebranes, and partly in a (compactified) $\bR^{1,1}$ along the branes.  In this way, one finds a ``supertube'' geometry that interpolates between the linear dilaton geometry of decoupled NS5-branes in the UV, and $AdS_3\times\bS^3$ in the IR.

In Section~\ref{sec:supertube}, we construct this family of gauged WZW models at $\nfive=1$.  The hybrid formulation using $\psu$ at level $\nfive=1$ provides a worldsheet description that precisely maps to the $1/N$ expansion of the spacetime dual CFT, which is the symmetric product orbifold $(\bT^4)^N/S_N$~\cite{Gaberdiel:2018rqv, Eberhardt:2018ouy,Eberhardt:2019ywk, Eberhardt:2020akk, Eberhardt:2020bgq, Dei:2020zui, Knighton:2020kuh, Gaberdiel:2021njm, Gaberdiel:2022als, Dei:2022pkr, Dei:2023ivl, McStay:2023thk, Knighton:2023mhq}.  
On the other hand, the interpolating geometry has been argued to be dual to the (single-trace) $T\bar T$ deformation of the spacetime CFT~\rcite{Giveon:2017nie, Asrat:2017tzd, Giveon:2017myj, Araujo:2018rho, Apolo:2019zai, Chang:2023kkq, Dei:2024sct}. The gauged WZW model provides in the $\nfive=1$ context an explicit, calculable model for holography beyond AdS/CFT.

While we only discuss in detail the gauged WZW formulation of a family of two-charge (NS5-F1) supertubes, a larger three-charge (NS5-F1-P) family is achieved by a further generalization of the gauge group embedding~\rcite{Martinec:2017ztd,Bufalini:2021ndn}.  This generalization is a straightforward extension of the null gauging formalism which will work as well for $\nfive=1$ as it does for $\nfive>1$.  The infrared geometries of all of these backgrounds is $(AdS_3\times\bS^3)/\bZ_\sfk$ for different choices of $\bZ_\sfk$ isometry, and should connect to the analysis of such orbifolds in~\rcite{Gaberdiel:2023dxt}.

Several appendices supply details and background material.  Appendix~\ref{sec:RNS review} reviews the RNS formulation of worldsheet string theory, and sets our notational conventions.  Details of the hybrid formulation and the RNS\,$\leftrightarrow$\,hybrid map are the subject of Appendix~\ref{sec:hybrid map}.   Details of $\psu_1$ representation theory are presented in Appendix~\ref{sec:details-repr-theory}, while the current algebra OPE's of $\psu$ for general level are the topic of Appendix~\ref{sec:algebras}.  Details of the null-gauged WZW model are the subject of Appendix~\ref{sec:null details}.

\vskip .7cm
Our work demonstrates the consistency of worldsheet string theory in the throat of a single NS5-brane, and elucidates its connection with the more traditional RNS worldsheet formalism.  Let us list several areas ripe for further investigation:
\begin{enumerate}
\item
While we have touched on a few aspects of the BRST cohomology at $\nfive=1$, a more complete and detailed analysis is in progress for both the pure fivebrane and the interpolating NS5-F1 supertube geometry, and will be reported elsewhere.  
\item
At low energies, the worldsheet theory provides a dual description of the 5+1d abelian gauge theory on the fivebrane.  How does the holographic map work in detail?  
\item
What is the full extent of the spacetime symmetry algebra?
For instance, there are charged states in the compactified fivebrane theory corresponding to wrapped D-branes bound to the fivebrane, discussed in~\rcite{Larsen:1999uk,Aharony:2024fid}.  It would be interesting to understand their description in our formalism.
\item
The worldsheet theory provides a way to perturbatively calculate correlation functions in the decoupled fivebrane theory.  Correlation functions and S-matrix elements in little string theory for $\nfive\ge2$ were computed in~\rcite{Aharony:2004xn,Chang:2014jta,Lin:2015zea}.  What are their analogues for $\nfive=1$?  The gauged WZW framework allows us to simply take the correlation functions of $\cG$ and select those that are invariant under $\cH$.
\item
The computation of observables in (single-trace) $T\bar T$ deformed CFTs should also be within reach.  
\item
A similar analysis should be possible for the large $\cN=4$ theory and its WZW model on the supergroup $D(2,1;\alpha)$~\rcite{Eberhardt:2019niq, Gaberdiel:2024dva, Eberhardt:2025sbi}, describing string theory in $AdS_3\times\bS^3\times\bS^3\times\bS^1$, as well as other backgrounds involving fivebranes, see \eg~\rcite{Brennan:2020bju}.
\item
While our focus in this work is on $\nfive=1$, a byproduct of our analysis is a description of $AdS_3$ string theory with non-vanishing R-R moduli using hybrid variables, which is valid for any $\nfive$.  It will be interesting to explore the effect of this deformation on the spectrum and dynamics of the worldsheet theory for $\nfive>1$ as well as $\nfive=1$.

The deformation of Section~\ref{sec:non-holo} is an example of the sort of asymptotic wall (growing toward the $AdS_3$ boundary) that appears in the analysis of Seiberg and Witten~\rcite{Seiberg:1999xz}.  When the background consists of only F1 and NS5 fluxes (with $\nfive>1$), and the R-R moduli are set to zero, there is a continuum of ``long string'' states in worldsheet string dynamics~\rcite{Giveon:1998ns,Maldacena:2000hw}.  The Seiberg-Witten wall lifts this continuum, but their analysis does not account for the ``AdS wall'' discussed in Section~\ref{sec:non-holo}.  It would be interesting to explore the string spectrum and dynamics in the presence of both walls for $\nfive>1$.  
\end{enumerate}

\bigskip
\bigskip

\section{The hybrid formalism for \texorpdfstring{$\boldsymbol{\nfive>1}$}{}}
\label{sec:hybrid kgt1}

We begin with a summary of worldsheet string theory on $AdS_3\times \bS^3\times \bT^4$ in the RNS formulation, and its transcription to hybrid variables.

\subsection{RNS summary}
\label{sec:RNS summary}

The RNS worldsheet formulation of type II string theory consists of a suitable superconformal field theory describing the target spacetime, with the superconformal symmetry gauged so that physical observables are independent of the worldsheet parametrization.  We will be interested in backgrounds of the form $AdS_3\times\bS^3\times \bT^4$.
The $AdS_3\times\bS^3$ factor is a group manifold, and the background electric and magnetic NS three-form flux sourced by the background fundamental strings and NS5-branes generate Wess-Zumino terms in the corresponding $\SLtwo$ and $\SUtwo$ nonlinear sigma models, respectively, turning them into Wess-Zumino-Witten models.  These worldsheet CFT's are completely characterized by their $\sltwo\times\sutwo$ current algebra symmetries, which we review in Appendix~\ref{sec:RNS review}.

In particular, the BRST charge that enforces superconformal invariance is
\begin{equation}
Q_\str = \oint\! dz \,\sfJ_\brst(z)  = \oint\! dz \Big[\cstr \Big(T_{\rm m} +\frac{1}{2}T^{\rm gh}_s \Big) - \gammastr \Big(  G_{\rm m} +\frac{1}{2}G^{\rm gh}_s \Big)\Big] ~, 
\label{BRST-RNS-intro}
\end{equation}
where $\cstr,\gammastr$ are (super)reparametrization ghosts, $T_{\rm m}, T^{\rm gh}_s$ are the matter and ghost stress tensors generating conformal transformations on the matter and ghost sectors, and $G_{\rm m}, G^{\rm gh}_s$ are the corresponding supercurrents generating superconformal transformations.  
These symmetry generators, and thus the BRST charge, can be written entirely in terms of the currents and their superpartners via the Sugawara construction.

The cohomology of the BRST charge defines the physical state space and observables.  Our primary focus here will be to construct a suitable BRST operator for string theory at $\nfive=1$.  We begin in this section by reviewing the formulation of string theory in Green-Schwarz style ``hybrid'' variables for $\nfive>1$, followed by its extension to $\nfive=1$ in the following section.

\subsection{The RNS \texorpdfstring{$\leftrightarrow$}{} hybrid map}
\label{sec:RNS2hybrid}

An alternative realization of worldsheet string theory on $AdS_3\times\bS^3\times\bT^4$ was introduced in~\rcite{Berkovits:1999im}.  It is called the hybrid formalism, because for some of the target spacetime it retains an RNS-like description with the worldsheet fermions living in the tangent bundle, while in the remainder of the target spacetime the worldsheet fermions live in a spinor bundle.  This reformulation allows some aspects of spacetime supersymmetry to be incorporated in a more manifest fashion.

In this subsection we review the relation between the RNS formalism and this hybrid formalism.  Unfortunately, by working simultaneously with two separate formalisms, as well as introducing eventually three separate components of the BRST operator, there will be a lot of notational overhead.  To help the reader keep track, we will notationally distinguish ghosts via the use of bold font, with a subscript $s$ to denote worldsheet superconformal ghosts (similarly, below we will use the subscript~$n$ to denote null gauging ghosts).  
We bosonize the RNS spinor ghosts as 
\begin{equation}
\label{ghbos}
\betastr = e^{-\phistr + \chistr } \partial \chistr ~, \qquad \gammastr = e^{\phistr- \chistr }  ~,
\end{equation}
where 
\begin{equation}
\label{ghbos 2pt}
\phistr(z) \phistr(w) \sim - \log(z-w) \,, \qquad  \chistr(z) \chistr(w) \sim \log(z-w) ~. 
\end{equation}
We also bosonize the ten RNS fermions as
\begin{align}
\partial H_1 &= \frac{1}{\nfive }(\psi^+ \psi^-) ~~, & \partial H_2 &= \frac{1}{\nfive }(\chi^+ \chi^-) ~~, & \partial H_3 &=-\frac{2}{\nfive }(\psi^3 \chi^3) ~~, \\ 
\partial H_4 &=(\zeta_1 \zeta_1^*) ~~, & \partial H_5 &= (\zeta_2  \zeta_2^*) ~~,  & & 
\end{align}
where the free bosons $H_i$ with $i=1, \dots 5$ have the OPE 
\begin{equation}
H_i(z) H_j(w) \sim \delta_{ij} \log(z-w) ~. 
\end{equation}
The RNS free fermions can be expressed in terms of the bosons $H_i$ as
\begin{equation}
\begin{aligned}
\psi^\pm &= \sqrt \nfive \, e^{\pm H_1} \,, &  \chi^\pm &= \sqrt \nfive \, e^{\pm H_2} \,, & \psi^3 \pm \chi^3 &= \sqrt \nfive \, e^{\pm H_3} \,,\quad & &  \\[.1cm]
\zeta_1 &= e^{H_4} \,, & \zeta_2 &= e^{H_5} \,,   &  \zeta_1^* &= e^{-H_4} \,, &  \zeta_2^* &= e^{-H_5} ~. 
\end{aligned}
\label{RNS-fermions-bosonization}
\end{equation}
Then we define the hybrid variables 
\begin{align}
\label{hybrid map}
p^{\alpha \beta} & = e^{\frac{\alpha}{2} H_1 + \frac{\beta}{2} H_2 - \frac{\alpha \beta}{2} H_3 - \frac{1}{2}H_4 - \frac{1}{2}H_5 -\frac{1}{2}\phistr } ~, \nn\\
\theta^{\alpha \beta} &= e^{\frac{\alpha}{2} H_1 + \frac{\beta}{2} H_2 + \frac{\alpha \beta}{2} H_3 + \frac{1}{2}H_4 + \frac{1}{2}H_5 + \frac{1}{2}\phistr } ~, \nn\\
\Psi_1 &= e^{H_4 + \phistr - \chistr} ~,\qquad
\Psi^*_1 = e^{-H_4 - \phistr + \chistr} ~,  \\
\Psi_2 &= e^{H_5 + \phistr - \chistr} ~,\qquad
\Psi^*_2 = e^{-H_5 - \phistr + \chistr} ~, \nn\\
\rhob & = -2 \phistr - H_4 - H_5 + \chistr ~, \nn
\end{align}
which obey the algebra\footnote{In our conventions $\vareps^{+-}=-\vareps_{+-} = 1$.}
\begin{align}
\label{hybrid OPEs}
p^{\alpha \beta}(z) \,  \theta^{\gamma \delta}(w) &\sim \frac{\vareps^{\alpha\gamma} \vareps^{\beta \delta}}{z-w} ~, 
\nn\\[.2cm]
\Psi^{~}_\sfi(z) \, \Psi^*_\sfj(w) & \sim \frac{\delta_{\sfi\sfj}}{z-w} \,,  \qquad \sfi, \sfj \in \{ 1 , 2 \} ~, 
\\[.3cm]
\rhob(z) \, \rhob(w) & \sim - \log(z-w) ~,  
\nn
\end{align}
respectively. All other OPE's are regular.

The hybrid variables provide an alternate orthonormal basis in the GSO-even field space \cite{Berkovits:1999im}, as we review in Appendix~\ref{sec:hybrid map}.  Any GSO-even operator in the RNS formalism can be translated to an equivalent expression in the hybrid formalism. 
Thus one has the following equivalence of stress tensors: 
\begin{equation}
T^{\text{RNS}} = \mathcal T^{\text{Hybrid}} ~, 
\end{equation}
see Appendices~\ref{sec:RNS review} and \ref{sec:hybrid map} for our conventions and definitions. 
Note that the stress tensors of the free fermions $p^{\alpha \beta},\theta^{\alpha \beta}$ as well as $\Psi^*_\sfj, \Psi^{~}_\sfj$ are ``topologically twisted'', so that these conjugate pairs of fields have conformal dimension $(1,0)$ instead of the worldsheet dimension 1/2 of the free fermions $\psi^a,\chi^a,\zeta^{~}_\sfj,\zeta^*_\sfj$ of the RNS formalism.

\subsection[\texorpdfstring{$\psu$}{}]{\texorpdfstring{${\boldsymbol{\psu}}$}{}}
\label{sec:psu}

The motivation for reformulating string theory in hybrid variables is to make some aspects of spacetime supersymmetry more manifest by replacing the worldsheet superconformal 
$\sltwo\times\sutwo$ current algebra by an ordinary current algebra on the supergroup $\psu$, which has the same bosonic subalgebra. Some of the spacetime supersymmetries are realized simply in terms of the primitive anticommuting worldsheet fields $p^{\alpha\beta}$ and their conjugates $\theta^{\alpha\beta}$ (of dimension $(1,0)$ respectively).
These transform as bispinors of $\sltwo\times\sutwo$ instead of the vector representation employed by the RNS formalism.

The free fermions \eqref{hybrid OPEs} generate a fermionic $\sltwo_{-2}\times\sutwo_2$ current algebra via 
\begin{align}
\label{fermcurr}
J_{\rm fer}^a = -\hf \tau^a_{\alpha\beta}\,\vareps_{\mu\nu} \, p^{\alpha\mu}\theta^{\beta\nu} ~,\qquad
K_{\rm fer}^a = -\hf \vareps_{\alpha\beta}\,\sigma^a_{\mu\nu} \, p^{\alpha\mu}\theta^{\beta\nu} ~,
\end{align}
where our group theory conventions are spelled out in Appendix~\ref{sec:RNS review}.
One then combines the fermions with a bosonic $\sltwo_{\nfive +2}\times\sutwo_{\nfive -2}$ algebra, so that the total current algebras 
\begin{align}
J^a &= J_{\rm bos}^a + J_{\rm fer}^a \,, & K^a &= K_{\rm bos}^a + K_{\rm fer}^a ~, 
\label{JK-n5bigger1}
\end{align}
both have level $\nfive$. The fermionic $\psu_{n_5}$ generators are then given by \cite{Bars:1990hx, Berkovits:1999im, Gotz:2006qp, Eberhardt:2019qcl}
\begin{subequations}
\label{Sgens}
\begin{align}
S^{\alpha\beta +} &= p^{\alpha\beta} ~, 
\label{Sgenplus}\\
S^{\alpha\beta -} &=  - (\tau_a)^{\alpha}_{~\gamma}(J^a_{\rm bos}+\hf J^a_{\rm fer})\theta^{\gamma\beta} + (\sigma_a)^{\beta}_{~\gamma}(K^a_{\rm bos}+\hf K^a_{\rm fer})\theta^{\alpha\gamma} - n_5 \,\partial\theta^{\alpha\beta} ~,
\label{Sgenminus}
\end{align}
\end{subequations}
see Appendix~\ref{sec:algebras} for our conventions.
The fact that the $p^{\alpha\beta}$ are some of the fermionic generators of $\psu$ allows the contribution of $\psu$ to the BRST charge to be written entirely in terms of its current algebra.

To complete the worldsheet theory, this worldsheet supergroup current algebra representation is combined with a topologically twisted CFT for the flat directions $X^{~}_\sfj, X^*_{\sfj}, \Psi^{~}_\sfj,\Psi^*_\sfj$ (see Appendix~\ref{sec:hybrid map}), together with a set of BRST ghosts $\bstr,\cstr$ and $\rhob$. 

The analysis of~\rcite{Berkovits:1999im} shows that string theory on $AdS_3\times\bS^3\times \bT^4$ (with purely NS fluxes) has a formulation in terms of the $\PSU_\nfive$ WZW model, together with these additional ingredients.  The map~\eqref{hybrid map} relates this formulation to the standard RNS formulation in terms of the worldsheet supersymmetric WZW model.


\subsection{The BRST charge}

One can use this RNS\,$\leftrightarrow$\,hybrid map to rewrite the BRST operator \eqref{BRST-RNS-intro} in hybrid variables~\rcite{Berkovits:1999im},  
\be
\label{10d Qhybrid}
Q_\str = \oint\! dz \, \mfJ_\brst(z) = \oint\! dz \, (\mfJ_0 + \mfJ_1 + \mfJ_2 )(z) ~,
\ee
where 
\begin{align}
\mfJ_0 &= \big(\cT_{\mathfrak{psu}} + \cT_{\flat}  + \cT_\rhob \big)\cstr + \partial \cstr (\bstr \cstr) - \cG_{\flat}^+ -\partial \big(\partial \rhob \, \cstr -\partial \cstr + 2 \cJ^{\mathcal R}_{\flat} \, \cstr  \big) ~, 
\nn \\[.2cm]
\mfJ_1 &= \frac{e^{-\rhob}}{2 \, \sqrt{\nfive}}\Big[ \big(\vareps_{\mu\nu} \, (\tau_a)_{\alpha\beta} \, J^a +  \vareps_{\alpha\beta} \, (\sigma_a)_{\mu\nu} \, K^a \big)p^{\alpha\mu}p^{\beta\nu} - 4 \, \vareps_{\alpha \beta} \, \vareps_{\mu \nu} \, p^{\alpha\mu}\partial p^{\beta\nu} \Big]    ~, 
\label{hyb-BRST-before-similarity}
\\[.2cm]
\mfJ_2 &= -e^{-2\rhob} p^{++} p^{+-} p^{-+} p^{--} \big( \cG_{\flat}^- + \bstr \big)  ~.  \nn
\end{align} 
Various ingredients such as stress tensors, \etc, entering eqs.~\eqref{hyb-BRST-before-similarity} are spelled out in Appendix~\ref{sec:hybrid map}, together with the specific choice of normal ordering. 
In particular, the $\bT^4$ supercurrent has been twisted and split into its $\cN=2$ components $\cG^+_{\flat}$ and $\cG^-_{\flat}$, which have dimension $(1,2)$ after mapping $\zeta_\sfj \to \Psi_\sfj$, $\zeta^*_\sfj \to \Psi^*_\sfj$ (see Appendix~\ref{sec:RNS review}). 
One can check that the BRST current $\mfJ_\brst$ has a regular OPE with itself. More broadly, the string BRST current $\mfJ_\brst = \mfJ_0+\mfJ_1+\mfJ_2$ belongs to a twisted $\cN=2$ superconformal algebra on the worldsheet~\rcite{Berkovits:1999im}. The other generators of this algebra are important ingredients in the calculation of correlation functions and are spelled out in Appendix~\ref{sec:hybrid map}.

The BRST charge can be simplified via a (conformally invariant) similarity transformation on its component operators 
\be
\label{simtransf}
\cO\to e^R\cO e^{-R}
~,\qquad
R = \oint\!dz\, \cstr \, \cG_{\flat}^-  ~.
\ee
The nilpotence of $R$ implies that the transformation of the 
BRST current involves only a small number of commutators with $R$, 
\begin{align}
 \mfJ_0 \to e^R \, \mfJ_0 \, e^{-R} &= \mfJ_0 + [R,\mfJ_0] + \tfrac12 [R,[R,\mfJ_0]] ~, \\
 \mfJ_1 \to e^R \, \mfJ_1 \, e^{-R} &= \mfJ_1 ~, \\
 \mfJ_2 \to e^R \, \mfJ_2 \, e^{-R} &= \mfJ_2 + [R,\mfJ_2]        ~. 
\end{align}
The transformation has the effect of removing $\cT_{\flat}$ from $\mfJ_0$ and $\cG^-_{\flat}$ from $\mfJ_2$, as well as modifying the ghost terms in $\mfJ_0$, so that the BRST current becomes \rcite{Berkovits:1999im, Gaberdiel:2022als} 
\begin{subequations}
\label{J hybrid}
\begin{align}
\mfJ_0 &= \Big[\cT_{\mathfrak{psu}} - \hf \partial(\rhob\tight+ i\sigmab) \partial(\rhob\tight+ i\sigmab) + \hf\partial^2(\rhob\tight+ i \sigmab) \Big] e^{i\sigmab} - \cG_{\flat}^+ ~, 
\label{J0 hybrid}\\[.2cm]
\mfJ_1 &= \frac{e^{-\rhob}}{2 \, \sqrt{\nfive}}\Big[ \big(\vareps_{\mu\nu} \, (\tau_a)_{\alpha\beta} \, J^a +  \vareps_{\alpha\beta} \, (\sigma_a)_{\mu\nu} \, K^a \big)p^{\alpha\mu}p^{\beta\nu} - 4 \, \vareps_{\alpha \beta} \, \vareps_{\mu \nu} \, p^{\alpha\mu}\partial p^{\beta\nu} \Big]  ~, 
\label{J1 hybrid}
\\[.2cm]
\mfJ_2 &= - e^{-2\rhob-i \sigmab} p^{++} p^{+-} p^{-+} p^{--}   ~, 
\label{J2 hybrid}
\end{align}
\end{subequations}
where $\partial\sigmab=i \bstr\cstr$ bosonizes the reparametrization ghosts, see eq.~\eqref{bscs-bosonization}, and additional details including our choice of normal ordering are provided in Appendix~\ref{sec:hybrid map}.

\bigskip
\section{The hybrid formalism for \texorpdfstring{$\boldsymbol{\nfive=1}$}{} }
\label{sec:levelone}

Conventional wisdom says that there is no worldsheet description of a background with a single NS5-brane in string theory.  The reasoning goes that the $\sutwo$ fermions $\chi^a$ of~\eqref{su2 alg} already form an $\sutwo$ current algebra at level $k_\fer=2$, and the total current algebra must have level $\nfive=k_\su+k_\fer$, so the underlying bosonic current algebra for a single fivebrane would have to have level $k_\su =-1$.  The representation theory for negative level has all sorts of negative norm states, and so the usual string theory no-ghost theorem would fail.

It thus came as somewhat of a surprise when a worldsheet description of string theory on $AdS_3\times\bS^3\times\bT^4$ at level $\nfive=1$ was proposed in~\rcite{Gaberdiel:2018rqv, Eberhardt:2018ouy}.  This construction is able to bypass the above no-go argument because a single pair of hybrid fermions, \eg\ $p^{+\beta},\theta^{-\beta}$, realize a current algebra at level one because they are $\sutwo$ spinors instead of $\sutwo$ vectors.

\subsection[The minimal realization of \texorpdfstring{$\psu_1$}{}]{The minimal realization of \texorpdfstring{$ \boldsymbol{\psu_1}$}{}}
\label{sec:min}

At level one, $\psu$ has the following ``minimal'' free field realization~\rcite{Beem:2023dub, Dei:2023ivl}
\begin{align}
\label{keq1}
J^+ &= \beta^+ ~~,
&K^+ &= -p^{++}\theta^{-+} ~, &
\nn\\
J^3\, &= \gamma^-\beta^+ -\hf \vareps_{\alpha\beta}\,p^{+\alpha}\theta^{-\beta}  
~~,  
&K^3\, &= \hf(p^{++}\theta^{--} + p^{+-}\theta^{-+})  &
\nn\\
J^- &= \gamma^-\big(\gamma^-\beta^+ - \vareps_{\alpha\beta}\,p^{+\alpha}\theta^{-\beta}\big) - \partial \gamma^- ~~,
&K^- &= p^{+-}\theta^{--} & \nn 
\\
S^{+++} &= p^{++} ~, \qquad~~~~
&S^{+-+} &= p^{+-} 
\\
S^{--+} &= -\gamma^- p^{+-} 
~~, \qquad
&S^{-++} &= -\gamma^- p^{++} ~~, &
\nn\\
S^{+--} &= -\beta^+\theta^{--} ~~,
&S^{++-} &= -\beta^+ \theta^{-+}  ~~,  &
\nn\\
S^{-+-} &= \big(\beta^+\gamma^- - \vareps_{\alpha\beta}\,p^{+\alpha}\theta^{-\beta}\big)\theta^{-+} ~~,
&S^{---} &= \big(\beta^+\gamma^- - \vareps_{\alpha\beta}\,p^{+\alpha}\theta^{-\beta}\big)\theta^{--} ~, &
\nn
\end{align}
where $(\beta^+, \gamma^-)$ is a Grassmann-even $\beta \gamma$ system of conformal weight (1,0) obeying the OPE 
\begin{subequations}
\label{minimal-free-fields}
\be 
\gamma^-(z)\, \beta^+(w) \sim \frac{1}{z-w} ~, 
\label{gammam-betap-OPE}
\ee
while the Grassmann-odd fields $(p^{++},\theta^{--})$, $(p^{+-},\theta^{-+})$ are $bc$ systems of conformal weight (1,0), with 
\be
p^{+\alpha}(z)\, \theta^{-\beta}(w) \sim \frac{\vareps^{\alpha \beta}}{z-w} ~. 
\ee
\end{subequations}
As mentioned above, the free fields $p^{\alpha\beta}$ appearing in the BRST current~\eqref{J hybrid} should be thought of as the fermionic currents $S^{\alpha\beta+}$ of~\eqref{Sgenplus}.  So making the replacement $p^{\alpha\beta}\to S^{\alpha\beta+}$ in~\eqref{J hybrid}, and then substituting the above expressions for these currents, one finds the BRST operator for this realization of level $\nfive=1$~\rcite{Dei:2023ivl}.%
\footnote{An alternative, equivalent free field realization using $\mathfrak{u}(1,1|2)$ was previously discussed in~\rcite{Gaberdiel:2022als}, together with a set of ghost fields to implement a gauging of $\mathfrak{u}(1,1|2)$ down to $\psu$.}

Note that the field content is substantially truncated from that of $\PSU_\nfive$ for level $\nfive>1$.  There, one has four independent free fermionic (1,0) systems 
$p^{\alpha\beta},\theta^{\alpha\beta}$ 
having total central charge $\mathtt c=-8$, together with an underlying bosonic $\SLtwo\times\SUtwo$ sigma model at levels $k_\sl=\nfive+2$, $k_\su=\nfive-2$, and combined central charge $\mathtt c\tight=6$.  Here, one has \emph{only two} free fermionic (1,0) systems $p^{+\alpha},\theta^{-\alpha}$ having combined central charge $\mathtt c\tight=-4$ (\ie\ not a complete $\sltwo\tight\times\sutwo$ bispinor multiplet), together with a single symplectic boson pair $\beta^+,\gamma^-$ of central charge $\mathtt c=2$.

Because of this mismatch, naively there is no longer an RNS equivalent of the $\nfive=1$ hybrid string~-- the number of fermion fields differs, and so there is no map that proceeds via bosonizing them and making a field redefinition.  To prepare the path for subsequent developments, in this section we construct a non-minimal realization of $\psu_1$ that {\it does} have an RNS equivalent.  Its shortcoming is that, because the bosonic $\sutwo$ current algebra has level $k_\su =-1$, the string naively fails to satisfy a no-ghost theorem.  To remedy this flaw, we add to the BRST charge an additional component $Q_\con$ that accomplishes two tasks: (1) The BRST cohomology of $Q_\con$ is isomorphic to the Hilbert space of the minimal realization of $\psu_1$ and hence equivalent to the original formulation of~\rcite{Eberhardt:2018ouy}; so that (2) the resulting worldsheet string theory is expected to satisfy a no-ghost theorem.  The extra constraint can be mapped back to an equivalent constraint in the RNS formulation, where it is a highly non-obvious constraint on the worldsheet currents associated to spacetime supersymmetry.

Let us now specify this non-minimal realization, as well as the additional constraint, and verify that it is equivalent to the minimal one we just reviewed, and compatible with the original string BRST constraints.

\subsection[A constrained non-minimal realization of \texorpdfstring{$\psu_1$}{}]{A constrained non-minimal realization of \texorpdfstring{$ \boldsymbol{\psu_1}$}{}}
\label{sec:nonmin}

In order to have an RNS\,$\leftrightarrow$\, hybrid map, one needs to augment the fermion content to a complete bispinor representation (just as one has for $\nfive>1$) by adding the missing components $p^{-\alpha},\theta^{+\alpha}$, so that we have four Grassmann-odd $bc$ systems with worldsheet dimensions $\Delta(p^{\alpha \beta})=1$, $\Delta(\theta^{\alpha \beta})=0$ for $\alpha, \beta \in \{+,-\}$,   obeying the OPE 
\begin{subequations}
\label{non-minimal-free-fields}
\be
p^{\alpha \beta}(z)\, \theta^{\gamma \delta}(w) \sim \frac{\vareps^{\alpha\gamma} \vareps^{\beta \delta}}{z-w} ~,  
\ee
and forming a full complement of bispinors of $\sltwo\times\sutwo$. 
Then, in addition to the minimal bosons
\be 
\gamma^-(z) \beta^+(w) \sim \frac{1}{z-w} ~, 
\ee
carrying $\mathfrak{sl}(2,\mathbb R)$ charges $J^3=\mp1$ as suggested by the notation, we introduce two additional Grassmann-even pairs $\betaup^{-\alpha},\gammaup^{+\alpha}$ of dimension $\Delta(\betaup^{-\alpha}) =1 $ and $\Delta(\gammaup^{+\alpha}) = 0$. 
The free fields $\betaup^{-\alpha},\gammaup^{+\alpha}$ have the same quantum numbers as the added fermions, but opposite statistics and are particular components of the bispinor representation of $\sltwo\times\sutwo$.  
We take the OPE of the additional bosons to be
\be
\gammaup^{+\alpha}(z) \betaup^{-\beta}(w)\, \sim \frac{\vareps^{\alpha\beta}}{z-w} ~.                                          
\label{lambda-beta-OPE}
\ee
\end{subequations}
From these, we form the currents
\begin{align}
J^+_\bos &= \;\beta^+ \,,  & K^+_\bos &= \gammaup^{++}\betaup^{-+} ~, \nn \\ 
J^3_\bos \, & = ~\gamma^-\beta^+ + \hf \vareps_{\alpha\beta}\,\gammaup^{+\alpha}\betaup^{-\beta} \,, & K^3_\bos \, &= - \hf(\gammaup^{++}\betaup^{--}+\gammaup^{+-}\betaup^{-+}) ~, \label{keq1alt} \\
J^-_\bos &= \gamma^-\big(\gamma^-\beta^+ + \vareps_{\alpha\beta}\,\gammaup^{+\alpha}\betaup^{-\beta}\big) - 3\partial\gamma^- \,,  & K^-_\bos & = -\gammaup^{+-}\betaup^{--} ~, \nn
\end{align}
which close into an $\sltwo$ algebra at level three, and an $\sutwo$ algebra at level minus one.%
\footnote{Note that the $\sutwo$ currents are simply $\sigma^a_{\mu\nu}\gammaup^{+\mu}\betaup^{-\nu}$, where the raised index on the usual Pauli matrices $(\sigma^a)_\mu{}^{\kappa}$ has been lowered with the spinor metric $\varepsilon_{\kappa\nu}$.} The $\mathfrak{su}(2)_{-1}$ algebra is realized in terms of two pairs of symplectic bosons as discussed in \cite{Gaberdiel:2018rqv}, while the extra $U(1)$ current $\vareps_{\alpha\beta}\,\gammaup^{+\alpha}\betaup^{-\beta}$ is used to construct the $\sltwo$ algebra at level $3$. 
Indeed, one can check that the Sugawara stress tensor built from the current algebra is the same as the free field stress tensor
\be
\label{same T}
T_{\sl}^\bos + T_{\su}^\bos = - \beta^+ \partial \gamma^- - \varepsilon_{\alpha\beta} \,  \betaup^{-\alpha} \partial \gammaup^{+\beta} ~, 
\ee
where the central charge of the bosonic $\sltwo$ stress tensor $T_{\sl}^\bos$ is $\mathtt c_{\sl}=\frac{3k_\sl}{k_\sl-2}=9$ for $k_\sl =3$ while the central charge of the bosonic $\sutwo$ stress tensor $T_{\sl}^\bos$ is $\mathtt c_{\su}=\frac{3k_\su}{k_\su+2}=-3$ for $k_\su =-1$. 
The bosonic current algebra~\eqref{keq1alt} also has Virasoro central charge six, being made of three pairs of symplectic bosons; the equivalence of stress tensors guarantees that this also matches the current algebra central charge for $\sltwo_3\times\sutwo_{-1}$.

The bosonic currents \eqref{keq1alt}, together with the fermionic currents~\eqref{fermcurr}, realize a $\psu$ algebra of level one. The $\sltwo_1 \oplus \sutwo_1$ is simply obtained as, 
\begin{subequations}
\label{keq1-nonmin-all}
\be
J^a = J^a_\bos + J^a_\fer \,, \qquad K^a = K^a_\bos + K^a_\fer ~,
\label{sl2-su2-nonminimal}
\ee
with the bosonic $\sltwo_3 \times \sutwo_{-1}$ currents $J^a_\bos, K^a_\bos$ given by eq.~\eqref{keq1alt} and the fermionic $\sltwo_{-2} \times \sutwo_2$ currents $J^a_\fer,K^a_\fer$ given in~\eqref{fermcurr}. Supercurrents are realized as in \eqref{Sgens} with $n_5 =1$, which we rewrite here for convenience, 
\begin{align}
\label{easy S}
S^{\alpha\beta +} &= p^{\alpha\beta} ~, \\
\label{hard S}
S^{\alpha\beta -} &=  - (\tau_a)^{\alpha}_{~\gamma}(J^a_{\rm bos}+\hf J^a_{\rm fer})\theta^{\gamma\beta} + (\sigma_a)^{\beta}_{~\gamma}(K^a_{\rm bos}+\hf K^a_{\rm fer})\theta^{\alpha\gamma} - \partial \theta^{\alpha\beta} ~, 
\end{align}
\end{subequations}
where again bosonic currents $J^a_\bos, K^a_\bos$ and fermionic currents $J^a_\fer,K^a_\fer$ are given in eqs.~\eqref{keq1alt} and \eqref{fermcurr} respectively. Indeed, making use of the free field OPE's \eqref{hybrid OPEs}, \eqref{gammam-betap-OPE} and \eqref{lambda-beta-OPE}, one can verify that the currents \eqref{keq1-nonmin-all} close into $\psu$ at level $n_5 =1$.\footnote{To carry out this and various other computations throughout this work, we made use of the Thielemans OPE package \cite{Thielemans:1991uw}.} See Appendix~\ref{sec:algebras} for our conventions. 

Notice that one has the same fermionic field content that appears for $\nfive>1$, see eqs.~\eqref{JK-n5bigger1} and \eqref{Sgens}, so there is still the map~\eqref{hybrid map} between the hybrid formalism and the standard RNS formalism.
While it is nice to have such a map, the negative level of the bosonic $\sutwo$ results in a non-unitary worldsheet theory.  On the other hand, there is no such issue with the minimal realization~\eqref{keq1}.

We can combine the advantages of both realizations by starting with the non-minimal realization and projecting out the unphysical content, leaving the minimal realization.
We can accomplish this reduction via the BRST constraint 
\begin{align}
\label{Qcon}
Q_\con = \oint\! dz \, \sfJ_\con(z) \, \qquad \text{with} \qquad \sfJ_\con = \vareps_{\mu\nu} \, \gammaup^{+\mu}  \big(\pup^{-\nu}\!+\gamma^- \pup^{+\nu}\big)  ~,
\end{align}
which relates half of the fermionic momenta to the other half, as is the case in~\eqref{keq1}. One can check that $Q_\con$ commutes with all the $\psu_1$ generators in \eqref{keq1-nonmin-all} and that 
\be
(Q_\con)^2 = 0 ~. 
\ee

The idea now is that within this non-minimal realization of $\psu_1$ there lies the minimal realization, up to $Q_\con$ exact terms.
Indeed, we have the following $Q_\con$ invariant (and not $Q_\con$ exact) operators
\begin{align}
\label{keq1 ops}
\gamma_\min^- \equiv \gamma^-
\,~~~&,~~~~~\,
\beta_\min^+ \equiv \beta^+ + \vareps_{\mu \nu} \pup^{+\mu}\thetaup^{+\nu} ~, 
\nn\\
p_\min^{+-} \equiv \pup^{+-}
~~&,~~~~~~
\theta_\min^{-+} \equiv \thetaup^{-+} + \gamma^- \thetaup^{++} ~, 
\\
p_\min^{++} \equiv \pup^{++}
~~&,~~~~~~
\theta_\min^{--} \equiv \thetaup^{--} + \gamma^- \thetaup^{+-} ~,
\nn
\end{align}
obeying 
\be 
\gamma^-_\min(z) \beta^+_\min(w) \sim \frac{1}{z-w} \,, \quad \quad p^{+\alpha}_\min(z) \theta^{-\beta}_\min(w) \sim \frac{\vareps^{\alpha \beta}}{z-w} ~, 
\ee 
while all the other OPE's are regular. The remaining fields we added decouple. In fact, of the remaining primitive free fields, $\gammaup^{+\mu}=-[Q_\con,\thetaup^{+\mu}]$ are BRST exact, while $\betaup^{-\mu}$ are not BRST invariant; the remaining two $\pup$'s are composites up to BRST exact quantities
\be
 \pup^{-\mu} = - \gamma^- \pup^{+\mu} - [Q_\con, \betaup^{-\mu}] ~,
\ee
while there aren't any other BRST invariant $\theta$'s besides $\theta^{-\mu}$.

Rewriting the non-minimal free field realization \eqref{keq1-nonmin-all} in terms of the fields defined in \eqref{keq1 ops} we obtain
\begin{subequations}
\be
\begin{aligned}
J^+ &= \beta^+_\min \\[.1cm]
J^3 & = \gamma^-_\min \beta^+_\min -\hf \vareps_{\alpha\beta}\,p^{+\alpha}_\min \, \theta^{-\beta}_\min + \tfrac12 \vareps_{\alpha \beta} \, Q_\con(\pi^{-\alpha} \theta^{+\beta}) ~,  \\[.1cm]
J^- &= \gamma^-_\min \big(\gamma^-_\min \beta^+_\min - \vareps_{\alpha\beta} \, p^{+\alpha}_\min \theta^{-\beta}_\min \big) - \partial \gamma^-_\min - \vareps_{\alpha \beta} \, Q_\con \bigl( \pi^{-\alpha} \theta^{-\beta} \bigr) ~, \\[.1cm]
K^+ &= -p^{++}_\min \theta^{-+}_\min - Q_\con \bigl(\pi^{-+} \theta^{++} \bigr) ~, \\[.1cm]
K^3 &= \hf(p^{++}_\min \theta^{--}_\min + p^{+-}_\min \theta^{-+}_\min) + \tfrac12 Q_\con \bigl( \pi^{--} \theta^{++} + \pi^{-+} \theta^{+-} \bigr)  ~, \\[.1cm]
K^- &= p^{+-}_\min \theta^{--}_\min + Q_\con \bigl( \pi^{--} \theta^{+-} \bigr) ~, 
\end{aligned}
\ee
and 
\be
\begin{aligned}
S^{+++} &= p^{++}_\min ~,  \\[.1cm]
S^{+-+} &= p^{+-}_\min ~,  \\[.1cm]
S^{--+} &= -\gamma^-_\min p^{+-}_\min ~, \\[.1cm] 
S^{-++} &= -\gamma^-_\min p^{++}_\min ~, \\[.1cm]
S^{+--} &= -\beta^+_\min \theta^{--}_\min + Q_\con \bigl( \pi^{--} \theta^{+-} \theta^{++} \bigr) ~, \\[.1cm]
S^{++-} &= -\beta^+_\min \theta^{-+}_\min + Q_\con \bigl( \pi^{-+} \theta^{+-} \theta^{++} \bigr) ~, \\[.1cm]
S^{-+-} &= \big(\beta^+_\min \gamma^-_\min - \vareps_{\alpha\beta} \, p^{+\alpha}_\min \theta^{-\beta}_\min \big) \theta^{-+}_\min - \vareps_{\alpha \beta} \, Q_\con \bigl( \pi^{-\alpha} \theta^{- \beta} \theta^{++} \bigr) ~, \\[.1cm]
S^{---} &= \big(\beta^+_\min \gamma^-_\min - \vareps_{\alpha\beta} \, p^{+\alpha}_\min \theta^{-\beta}_\min \big) \theta^{--}_\min - \vareps_{\alpha \beta} \, Q_\con \bigl( \pi^{-\alpha} \theta^{- \beta} \theta^{+-} \bigr) ~. 
\end{aligned}
\ee
\end{subequations}
Comparing with eq.~\eqref{keq1} we see that, up to $Q_\con$ exact terms, we land exactly on the minimal free field realization. 

When we embed the constraint~\eqref{Qcon} in string theory, we must ensure compatibility with the string BRST operator.  One can show that the string BRST operator $Q_\str$~\eqref{10d Qhybrid} anticommutes with~$Q_\con$,
\be
\big\{Q_\str,Q_\con\big\} = 0 ~,
\ee
so that $(Q_\str+Q_\con)$ squares to zero.  Specifically, the vanishing of $\{ Q_0, Q_\con\}=0$ simply follows from $\sfJ_\con$ being a primary field of conformal dimension one under $\cT_{\mathfrak{psu}}$, and commuting with $\cG^+_{\flat}$. It is straightforward to check that $\{ Q_2, Q_\con\}=0$; all of the fields entering the definition of $\mfJ_2$, see eq.~\eqref{J2 hybrid}, have vanishing OPE with $\sfJ_\con$. Finally, the vanishing of the remaining term $\{ Q_1, Q_\con\}=0$ is simply a consequence of the regularity of the OPE's of $\sfJ_\con$ with $\rhob$, $p^{\alpha \beta}$, $K^a$ and~$J^a$.

Thus, the non-minimal realization of the $\nfive=1$ hybrid string, subject to the additional constraint $Q_\con$, is equivalent to the minimal realization that has been shown to match the perturbative behavior of the symmetric orbifold $(\bT^4)^N/S_N$ in the $1/N$ expansion \rcite{Eberhardt:2018ouy,Eberhardt:2019ywk, Dei:2023ivl}.

\subsection{The null vector decoupling the long string continuum}
\label{sec:nullvec}

A characteristic feature of the $n_5=1$ string is the absence of a long string continuum: The only representation entering the spectrum has $\mathfrak{sl}(2,\mathbb R)$-spin $j_\sl=\frac12$ \cite{Eberhardt:2018ouy}. This is a consequence of a shortening condition, taking the form of a null vector\footnote{See \cite{Eberhardt:2018ouy} for a detailed discussion of the representation theory of the model.}
\be
\mathcal N = T_{\mathfrak{psu}} - T^\bos_{\sl} - T^\bos_{\su} ~, 
\label{null-vector}
\ee
where $T_{\mathfrak{psu}}$ is the $\psu$ stress tensor at level $n_5 =1$, see eq.~\eqref{psu-sugawara-stress-tensor}, while $T^\bos_{\sl}$ and $T^\bos_{\su}$ are the $\sltwo$ and $\sutwo$ bosonic stress tensors at level $k_{\sl} = 1$ and $k_{\su} =1$, respectively. See eqs.~\eqref{su-sugawara-T} and \eqref{sl-sugawara-T} for their expression in terms of currents. In fact, one can check that upon substituting the minimal free-field realization \eqref{keq1} into \eqref{null-vector}, the vector on the left-hand-side vanishes identically; the decoupling of the null vector from the Verma module is automatically implemented in terms of free fields. 

One would then expect that also in terms of the non-minimal free field realization \eqref{keq1-nonmin-all} at level $n_5=1$, the vector \eqref{null-vector} and its descendants are decoupled from the spectrum. This is in fact the case, since the null vector \eqref{null-vector} is $Q_\con$-exact:
\be
\mathcal N = Q_\con \Sigma ~, 
\label{null-vector-Qcon-exact}
\ee
where the explicit form of $\Sigma$ can be found in Appendix~\ref{sec:the-null-vector-app}, see eq.~\eqref{Sigma}. In the next section we analyze the representation theory of the non-minimal formulation of the model and show that once $Q_\con$-exact states are decoupled from the spectrum, the only representations entering the spectrum are the ones identified in \cite{Eberhardt:2018ouy}, decoupling the long string continuum.

\subsection{Representation theory}
\label{sec:representation-theory}

In \cite{Eberhardt:2018ouy} it was shown that the $\psu_{n_5}$ affine algebra at level $n_5 =1$ only contains short representations, taking the form 
\begin{equation} 
\begin{tabular}{ccc}  & $(\mathcal C_\lambda^{\frac{1}{2}}, \boldsymbol 2)$ & \\
$(\mathcal C_{\lambda+\frac{1}{2}}^{1}, \boldsymbol 1)$ & & $(\mathcal C_{\lambda+\frac{1}{2}}^0, \boldsymbol 1)$
\end{tabular}
\label{short-rep}
\end{equation}
where $\mathcal C_\lambda^{j}$ denotes a continuous $\mathfrak{sl}(2,\mathbb R)$ representation with spin $j$ and Cartan eigenvalue $m$ with fractional part $m \in \mathbb Z + \lambda + \tfrac{1}{2}$, while $\boldsymbol 2$ and $\boldsymbol 1$ are respectively the spin $\tfrac12$ and the vacuum representation of $\sutwo$. The decoupling of the long string continuum is due to the presence of the null vector \eqref{null-vector}, imposing a shortening condition on the representations \cite{Eberhardt:2018ouy}. 

\paragraph{Representations in terms of the minimal free fields:}
When expressed in terms of the minimal free fields \eqref{minimal-free-fields}, the representations \eqref{short-rep} read \cite{Dei:2023ivl}
\be
\ket{m} \in (\mathcal C^0_{\lambda + \frac{1}{2}}, \boldsymbol 1) \,, \qquad \theta^{-\alpha}_0 \ket{m} \in (\mathcal C^{\frac{1}{2}}_{\lambda}, \boldsymbol 2) \,, \qquad  \theta^{-+}_0 \theta^{--}_0 \ket{m} \in (\mathcal C^{\frac{1}{2}}_{\lambda}, \boldsymbol 2) ~, 
\label{min-reps}
\ee
where the states $\ket{m}$ are defined as 
\be 
\beta_0^+ \ket{m} = m \ket{m+1} \,, \qquad \gamma^-_0 \ket{m} = \ket{m-1} \,, \qquad p^{+\alpha}_0 \ket{m} = 0 ~. 
\label{betagamma-reps}
\ee

\paragraph{Representations in terms of the non-minimal free fields:}
We showed in the previous section that when expressed in terms of the non-minimal variables \eqref{non-minimal-free-fields}, the null vector \eqref{null-vector} is $Q_\con$-exact, see eq.~\eqref{null-vector-Qcon-exact}. Hence, decoupling $Q_\con$-exact terms and restricting to the $Q_\con$ cohomology, one would expect to find the same representations identified in \cite{Eberhardt:2018ouy}, given in eq.~\eqref{short-rep}. In the following, explicitly constructing irreducible representations of the non-minimal zero mode algebra 
\be
[\beta^+_0, \gamma^-_0] = -1 \,, \qquad [\gammaup^{+\alpha}_0, \betaup^{-\beta}_0] \, = \vareps^{\alpha\beta} \,, \qquad \{ p^{\alpha \beta}_0, \theta^{\gamma \delta}_0 \} = \vareps^{\alpha \gamma} \vareps^{\beta \delta} ~, 
\label{zero-mode-algebra}
\ee
we show that this is indeed the case: The irreducible representations of the constrained non-minimal free field realization of $\psu_1$ are in one to one correspondence with the representations identified in \cite{Eberhardt:2018ouy}. We relegate some technical derivations to Appendix~\ref{sec:details-repr-theory}.

Representations of the non-minimal free fields \eqref{non-minimal-free-fields} can be constructed by tensoring three copies of representations of the form \eqref{betagamma-reps}. More precisely, for 
\be 
m_i \in \mathbb Z + \lambda_i + \frac12  \,, \qquad  \qquad  i = 1, 2, 3 ~, 
\ee
we can define the states $\ket{m_1, m_2, m_3}$ by the action of the zero modes~\eqref{zero-mode-algebra} as
\begin{subequations}
\label{zero-mode-rep}
\begin{align}
\beta_0^+ \ket{m_1, m_2, m_3} &= m_1 \ket{m_1+1, m_2, m_3} \,, & \gamma_0^- \ket{m_1, m_2, m_3}  &= \ket{m_1-1, m_2, m_3} ~, \\
\pi^{-+}_0 \ket{m_1, m_2, m_3} &= m_2 \ket{m_1, m_2+1, m_3} \,, & \lambda^{+-}_0 \ket{m_1, m_2, m_3} &= - \ket{m_1, m_2-1, m_3} ~, \\
\pi^{--}_0 \ket{m_1, m_2, m_3} &= m_3 \ket{m_1, m_2, m_3+1} \,, & \lambda^{++}_0 \ket{m_1, m_2, m_3} &= \ket{m_1, m_2, m_3-1} ~,
\end{align}
and 
\be 
p_0^{\alpha \beta}\ket{m_1, m_2, m_3} =0 ~. 
\ee
\end{subequations}
Notice that when some of the $\lambda_i$ take value $\lambda_i = \frac{1}{2}$, the `continuous' representations \eqref{zero-mode-rep} truncate and only positive integer values of $m_i$ appear, $m_i \in \mathbb Z_{\geq 0}$. This phenomenon was analyzed in detail in~\cite{Eberhardt:2018ouy}. While, once the dust settles, this does not play a crucial role when expressing representations in terms of minimal variables, in the following it will be important to distinguish whether $\lambda_2= \frac12$ and/or $\lambda_3 = \frac12$ and accordingly whether the representation truncates.\footnote{We will see in a moment that, in the same way that $\lambda = \frac12$ did not affect the representation theory analysis in terms of the minimal variables, it will not be important to distinguish whether $\lambda_1 = \frac12$ or not.} 

We show in Appendix~\ref{sec:details-repr-theory} that when $\lambda_2 \neq \frac12$ or $\lambda_3 \neq \frac12$ all the states in the representations \eqref{zero-mode-rep} that are $Q_\con$-closed are also $Q_\con$-exact. Equivalently, none of the states \eqref{zero-mode-rep} enter the $Q_\con$-cohomology when $\lambda_2 \neq \frac12$ or $\lambda_3 \neq \frac12$. In the same appendix, we also show that when instead $\lambda_2 = \lambda_3 =\frac12$, the only $Q_\con$-closed states that are not also exact, are 
\be
\ket{m_1, 0, 0} \,, \qquad \quad (\theta^{-\alpha}_\min)_0 \ket{m_1,0,0} \,, \qquad \quad (\theta^{-+}_\min)_0 \, (\theta^{--}_\min)_0 \ket{m_1,0,0} ~,
\label{non-min-reps}
\ee
where we remind the reader that the fields $\theta^{-\alpha}_\min$ have been defined in eqs.~\eqref{keq1 ops}. 
The states \eqref{non-min-reps} are clearly in one to one correspondence with the representations obtained in \eqref{min-reps} in terms of the minimal fields. In particular, as expected they reproduce the short representations~\eqref{short-rep}.

\subsection{RNS realization}
\label{sec:n5=1 RNS}

Due to the existence of the RNS\,$\leftrightarrow$\,hybrid map, the entire discussion above of $AdS_3\times\bS^3$ string theory at $\nfive=1$ could have been carried out in the RNS formalism.  For instance, the additional term $Q_\con$ in the BRST constraint can be written as
\be
\label{Qcon RNS}
Q_\con = \oint\!  dz \Big(\lambda^{+-} \big(e^{\sfv_3\cdot\sfH}-\gamma^- e^{\sfv_1\cdot\sfH}\big) +\lambda^{++}\big( e^{\sfv_4\cdot\sfH}-\gamma^- e^{\sfv_2\cdot\sfH}\big)\Big) ~,
\ee
where $\lambda^{+\mu}$ are primary fields in the bosonic $\SLtwo_3\times\SUtwo_{-1}$ WZW model (which we discuss more fully in Section~\ref{sec:non-holo} below),
and the RNS spin fields are given by 
\be
e^{-\phistr/2}\Sigma^{\alpha\mu\mfa} = {\exp}[ \sfv\tight\cdot  \sfH]
~~,~~~~
\sfv=\hf(\alpha,\mu,-\alpha\mu,\mfa,\mfa,-1,0) 
~~,~~~~
\sfH = (H_1,H_2,H_3,H_4,H_5,\phib_s,\chib_s)
~,
\ee
with the particular choices $\sfv_i$, $i=1,2,3,4$ given in Appendix~\ref{sec:hybrid map}.
Here $\alpha$ is the $\sltwo$ spin, $\mu$ is the $\sutwo$ spin, and $\mfa$ is the spinor polarization on $\bT^4$. 

From the perspective of the RNS formalism, the constraint~\eqref{Qcon} sets to zero a linear combination of spin fields.  Although it would be difficult to motivate such a modification, one could simply have postulated it, and proceeded without introducing all the machinery of the hybrid formulation.  The advantage of the hybrid formalism is that it expresses the constraint as a simple relation between elementary free fields, rather than a complicated relation among vertex operators.  

These spin fields are particular polarizations of spacetime supersymmetry generators, and thus eq.~\eqref{Qcon RNS} looks rather peculiar.  Remarkably, this procedure doesn't break any of the supersymmetries~-- they are reconstituted in terms of physical degrees of freedom in the BRST cohomology.  We will see why in Section~\ref{sec:non-holo}.

It has often been argued that there is no RNS worldsheet formulation of the throat of a single, isolated NS5-brane, precisely because of the negative level it would require for the bosonic $\SUtwo$ WZW model.  While there is indeed such a nonunitary component in our construction, we manage to bypass any resulting problems because additional constraints on the dynamics can be consistently imposed, which eliminate any negative norm states from the BRST cohomology beyond those of the usual unphysical string modes.  Although we have not proved it, we are confident that our construction satisfies a no-ghost theorem.  We showed that the non-minimal realization introduced above is equivalent to the previous constructions of the tensionless string \cite{Eberhardt:2018ouy, Dei:2023ivl}, and hence dual to the symmetric product orbifold of $\bT^4$, which is itself unitary.

\bigskip
\section{
R-R moduli in \texorpdfstring{$\boldsymbol{AdS_3\times\bS^3\times\bT^4$}}{} string theory  
}
\label{sec:PSU genl}

Physical string states for $\nfive=1$ are all based on the $j_\sl=1/2$ representation of $\sltwo$, whose dominant support is in the asymptotic region of $AdS_3$ \cite{Eberhardt:2018ouy, Eberhardt:2019ywk, Eberhardt:2020bgq, Eberhardt:2021jvj}.  In this section we examine the effect of turning on some of the four R-R moduli of string theory on this background, which has a strong effect on strings near the conformal boundary~\rcite{Seiberg:1999xz}; we will see that the form of the deformation is related to the constraint \eqref{Qcon} that we wish to impose for $\nfive=1$.  Indeed, we will argue that the imposition of the BRST constraint $Q_\con=0$ might be thought of in terms of the background having a nonzero value of a particular R-R modulus.  This idea illuminates the puzzles regarding the use of free field theory that were raised in the introduction, which afflict our analysis above.

Consider BPS R-R vertex operators in the RNS formalism.  They take the rough form 
\begin{align}
e^{-\half\phistr-\half\phistrbar}\, \big(\Sigma\bar \Sigma\,\Phi_{j_\sl}\Psi_{j_\su}\big)  ~,
\end{align}
where $\Phi_{j_\sl}, \Psi_{j_\su}$ are primary operators of the bosonic $\sltwo$ and $\sutwo$ current algebra CFT, respectively, and $\Sigma,\bar\Sigma$ are spin fields associated to the RNS fermions~\rcite{Friedan:1985ge}. 
Of particular interest is the choice $j_\sl=-1/2$, $j_\su=1/2$.  
The waveform $\Phi_{-1/2}$ is non-normalizable, as befits the worldsheet representative of a local operator in the dual spacetime CFT.
The spin fields transform in the $(-\hf,\hf)$ representation of $\sltwo\times\sutwo$.  One can then form the  $\sltwo\times\sutwo$ singlets
\be
\label{massterm}
\cM^{\mfa\bar\mfa} = e^{-\half\phistr-\half\phistrbar}\, \Sigma^{\alpha\mu\mfa}\bar \Sigma^{\beta\nu\bar\mfa}\,\Phi_{\alpha\beta}[g_\sl]\Psi_{\mu\nu}[g_\su]  ~,
\ee
which being $\sltwo\times\sutwo$ invariant, constitute a quartet of R-R moduli deformations of the background.  The singlet combination $\vareps_{\mfa\bar\mfa}\cM^{\mfa\bar\mfa}$ is the combination $C_0+C_4/V_4$ of the R-R scalar and four-form along $\bT^4$, while the other three combinations $\sigma^i_{\mfa\bar\mfa}\cM^{\mfa\bar\mfa}$ comprise the self-dual R-R two-forms $C_2^+$ on $\bT^4$. 
The operators~\eqref{massterm} are the representatives of the R-R moduli in the $(-1/2,-1/2)$ picture; one also needs their counterparts in the $(+1/2,+1/2)$ picture 
\begin{align}
\widetilde\cM^{\mfa\bar\mfa} = \big[Q_\str,\big\{\bar Q_\str, \xib_s\bar\xib_s\cM^{\mfa\bar\mfa}\big\} \big]
\end{align}
to deform the theory a finite distance along the moduli space.

We can be somewhat more explicit using the Wakimoto parametrization of the two-dimensional representation of $\SLtwo$
\be
\label{waki gsl}
\Phi[g_\sl] = \bigg(\begin{matrix} 1&~\bar\gamma\\0&~1\end{matrix}\bigg)\cdot \bigg(\begin{matrix} e^{-\frac 12\varphi_\sl}&~0\\0&~e^{+\frac 12\varphi_\sl}\end{matrix}\bigg)\cdot \bigg(\begin{matrix} 1&~0\\\gamma&~1\end{matrix}\bigg)  ~.
\ee
Then using the RNS~$\to$~hybrid map, the particular polarization $\mfa=\bar\mfa=+$ can be written directly in terms of $p,\bar p$ as%
\begin{align}
\label{ppbar term}
\tr\big[ g_\sl \,p\, g_\su\,\bar p \big] &=
e^{+\frac 12\varphi_\sl}\big( \Psi[g_\su]_{\mu\nu} \,\cC^\mu\bar\cC^\nu  \big) + e^{-\frac 12\varphi_\sl} \big(\Psi[g_\su]_{\mu\nu}\,p^{+\mu}\bar p^{+\nu}\big)  ~,
\end{align}
where
\be
\label{fermcon}
\cC^\mu = p^{-\mu} + \gamma^- p^{+\mu}
~,\qquad
\bar\cC^\mu = \bar p^{\,-\mu} + \bar\gamma^- \bar p^{\,+\mu}
\ee
are the constraints~\eqref{Qcon}.
We note that the interaction \eqref{ppbar term} between the fermions $p,\bar p$ and the bosonic WZW model fields was found in an analysis~\rcite{Gotz:2006qp} of the action of the $\PSU$ WZW model.

This interaction is distinct from the R-R modulus which is usually turned on in order to engineer string joining/splitting interactions, and that is related to the amount of R-R flux in the background; this latter deformation is the $\bT^4$ scalar singlet combination $\vareps_{\mfa\bar\mfa}\cM^{\mfa\bar\mfa}$, as opposed to the nilpotent deformation $\cM^{++}$ that is~\eqref{ppbar term} above.

Since the leading term in the expression is growing without bound in the asymptotic region (because the spin $j_\sl=-1/2$ representation corresponds to a non-normalizable operator), finite action physical states that are predominantly supported in the asymptotic region are those that satisfy the constraints $\cC^\mu=0$, $\bar \cC^\nu=0$.  For $\nfive=1$ these are the only physical states.

The above analysis suggests that the $PSU(1,1|2)$ WZW theory at level $\nfive=1$ necessarily has the R-R modulus~\eqref{ppbar term} turned on in the background. The effect of this modulus on the spectrum is however expected to be minimal \cite{Quella:2007hr}. In fact, the R-R vertex operator \eqref{ppbar term} is nilpotent and in absence of its charge conjugate $\mathcal M^{--}$, it has no impact on the $AdS_3$ or $\bS^3$ radii \cite{OhlssonSax:2018hgc} and hence does not affect the spectrum, in agreement with the analysis of \cite{Gaberdiel:2011vf, Gerigk:2012cq}.

\subsection[The effect of R-R moduli for
\texorpdfstring{$\nfive=1$}{}]{
The effect of R-R moduli for
\texorpdfstring{$\boldsymbol{\nfive=1}$}{}
}
\label{sec:non-holo}

The $\SLtwo$ WZW model is not holomorphically factorized into a product of free fields; rather, the worldsheet fields are asymptotically free at large $\varphi_\sl$ (much like Liouville theory), but an ``$AdS_3$ wall'' spoils their holomorphy away from the asymptotic boundary.  In the Wakimoto parametrization~\eqref{waki gsl}, the action of the bosonic $\SLtwo$ WZW model is given by
\be
\label{waki}
\cS^\SLtwo_{k_\sl}[g_\sl] = \frac{1}{4\pi}\int\! d^2z \Big[\beta^+\bar\partial\gamma^- + \bar\beta^+\partial\bar\gamma^- +\big(\coeff{k_\sl-2}4\big)\partial\varphi_\sl\bar\partial\varphi_\sl - \beta^+\bar\beta^+e^{-\varphi_\sl} \Big]
~.
\ee
The term $\beta^+\bar\beta^+e^{-\varphi_\sl}$ means that the dynamics at large $\varphi_\sl$ is only approximately free, in the limit of asymptotically large $\varphi_\sl$.

Previous work on $\nfive=1$, as well as our analysis above, has assumed that one could work with a free field realization of $\psu_1$ current algebra, and that this fully characterized the theory.  But here we have embedded the minimal realization of $\psu_1$ in a larger theory which has an underlying bosonic $\sltwo$ current algebra at level $k_\sl=\nfive+2=3$.  The $\SLtwo$ WZW model at level three is not a free theory.%
\footnote{The recent work~\rcite{Eberhardt:2025sbi} argues for the existence of a distinct theory with bosonic $\sltwo$ symmetry at level $k_\sl=3$, which is not the WZW model.}

We are claiming that the level one $\sltwo$ theory sits inside this larger WZW theory as a subset of BRST invariant states, but the $k_\sl=3$ bosonic WZW model needs an ``$AdS_3$ wall'' in order to be well-defined.  Such a wall breaks the holomorphy of various fields needed to build the BRST operator $Q_\con$, so one can and should worry about the consistency of the theory we are attempting to define.  We need holomorphy to write the constraint BRST operator, but the theory we are trying to write it in is not holomorphically factorized. The analysis above suggests that the additional constraint $Q_\con$ of~\eqref{Qcon} arises from a R-R deformation of the background, comparing~\eqref{Qcon} and~\eqref{fermcon}.  Indeed, such a term {\it does} appear in the $\PSU$ WZW model~\rcite{Gotz:2006qp}.  But this term is blowing up in the region of large $\varphi_\sl$, so once again one might worry that the theory is not free.

In this subsection, we address this issue by rewriting the field content in a way that isolates the $\SLtwo_3$ WZW model, and in so doing motivate the appearance of the constraint~\eqref{Qcon} not as an exact property of the $\PSU$ WZW theory but as an asymptotic approximation to it. The full WZW theory does not rely on the holomorphy of fields such as $\lambda^{+\mu}$ that appear in the BRST operator. Indeed, the BRST operator $Q_\con$ itself is only an asymptotic approximation to the underlying dynamics.

The $\sltwo$ and $\sutwo$ contributions to the symplectic bosons $\betaup^{-\mu},\gammaup^{+\mu}$ of~\eqref{lambda-beta-OPE} can be disentangled via bosonization analogous to~\eqref{ghbos}, \eqref{ghbos 2pt}
\be
\label{lampi bos}
\gammaup^{+\pm} = e^{\varphi_\pm}\eta_\pm = e^{\varphi_\pm - \varkappa_\pm} ~, \qquad
\betaup^{-\mp} = e^{-\varphi_\pm}\partial\xi_\mp = e^{-\varphi_\pm + \varkappa_\pm}\partial\varkappa_\pm ~, 
\ee
where in our conventions
\be 
\eta_\alpha(z) \xi_{\beta}(w) \sim \frac{\varepsilon_{\alpha \beta}}{z-w} ~. 
\ee
One has the Cartan subalgebra currents
\be
\label{JK3bos}
\begin{aligned}
J^3_\bos &= \gamma^-\beta^+ - \partial\varphi_\sl ~, &
\varphi_\sl &= \tfrac12(\varphi_+ + \varphi_-) ~,  \\
K^3_\bos &= -\partial\varphi_\su ~, &
\varphi_\su &= \tfrac12(\varphi_+ - \varphi_-) ~, 
\end{aligned}
\ee
as well as the raising/lowering operators
\be
\label{JKpmbos}
\begin{aligned}
J^+_\bos &= \beta^+
~,\qquad
& K^+_\bos &= - e^{+2\varphi_\su} \eta_+\partial\xi_+  = e^{+2\varphi_\su-\varkappa_++\varkappa_-}\partial\varkappa_-  ~, & \\
J^-_\bos &= \gamma^-(\gamma^-\beta^+ - 2\partial\varphi_\sl ) -3\partial \gamma^-
~,
& K^-_\bos &= e^{-2\varphi_\su} \eta_-\partial\xi_- = e^{-2\varphi_\su+\varkappa_+-\varkappa_-}\partial\varkappa_+  ~.&
\end{aligned}
\ee
Note that the bosons $\varphi_\sl$ and $\varphi_\su$ are normalized as 
\begin{equation}
    \partial\varphi_\sl(z) \partial\varphi_\sl(w) \sim -\frac{1}{2(z-w)^2} ~, \qquad \partial\varphi_\su(z) \partial\varphi_\su(w) \sim -\frac{1}{2(z-w)^2} ~. 
    \label{varphi-sl-su-OPEs}
\end{equation}
The boson $\varphi_\sl$ has an improvement term in its stress tensor, while $\varphi_\su$ does not
\be
T_{\varphi_\sl} = -(\partial\varphi_\sl)^2 - \partial^2\varphi_\sl ~,\qquad
T_{\varphi_\su} = - (\partial\varphi_\su)^2 ~. 
\ee
In these variables, one has a Wakimoto representation of $\sltwo_3$ in terms of $\beta^+,\gamma^-$ and $\varphi_\sl$ (see for instance~\rcite{Giveon:2001up} for a review), and a decoupled free field realization of $\sutwo_{-1}$ in terms of the timelike free scalar $\varphi_\su$ and the pair of (1,0) fermion systems $(\eta_\pm,\xi_\mp)$.  

Let us pause briefly to discuss this $\sutwo_{-1}$ theory. Formally, spin $j_\su$ representations of $\sutwo_{-1}$ have conformal dimension $\Delta=j_\su(j_\su+1)$, and have the free field realization 
\be
\label{Phisu}
\Phi^\su_{j_\su,m} = e^{2m\varphi_\su} \, e^{-(j_\su+m)\varkappa_+ - (j_\su-m)\varkappa_-} ~.
\ee
Their operator products with the currents are
\begin{align}
\label{KPhi OPE}
K^3(z) \, \Phi^\su_{j_\su,m}(w) &\sim \frac{m\,\Phi^\su_{j_\su,m}(w)}{z-w}  \nn\\[.2cm]
K^+(z) \, \Phi^\su_{j_\su,m}(w) &\sim \frac{(j_\su-m)\,\Phi^\su_{j_\su,m+1}(w)}{z-w}  \\[.2cm]
K^-(z) \, \Phi^\su_{j_\su, m}(w) &\sim \frac{(j_\su+m)\,\Phi^\su_{j_\su,m-1}(w)}{z-w} ~. \nn
\end{align}
One can in principle have any $j_\su\in\half\bZ$.  For $j_\su\ge0$, $\Phi_{j_\su,m}$ are the $2j_\su+1$ elements of the spin $j_\su$ representation of $\sutwo$ due to the null vectors at $m=\pm j_\su$ implied by the above OPE's; for $j_\su<0$, one has one-sided representations equivalent to the representations $D^\pm_{-j_\su}$ of $\sltwo$. 
This result agrees with the recent analysis of~\rcite{Mazzucchelli:2024jbk}. 
Note in particular that the $j_\su=\hf$ representation consists~of
\be
\Phi^\su_{\half,+\half} = e^{\varphi_\su-\varkappa_+}
~,\qquad
\Phi^\su_{\half,-\half} = e^{-\varphi_\su-\varkappa_-}  ~,
\ee
which when combined with the appropriate spin $j_\sl=-1/2$ operator (which we have heretofore been regarding as the free-field exponential $e^{\varphi_\sl}$) reconstitute the operators $\lambda^{+\mu}$ of eq.~\eqref{lampi bos}.

We now turn to the $\sltwo_3$ component.  In the usual Wakimoto realization of $\SLtwo$ conformal field theory, the fields are only approximately free at large $\varphi_\sl$, because the $AdS_3$ metric in Poincar\'{e} coordinates from which it arises has a $\varphi_\sl$-dependent interaction~\eqref{waki}.
This spoils the holomorphy of $\gammaup^{+\mu}$ and $\gamma^-$, which appear in the constraint $Q_\con$.%
\footnote{Also, the holomorphy of $\gamma^-$ is key to the construction of the holographic map to the dual symmetric product orbifold~\rcite{Eberhardt:2019ywk}.}
As a result, it appears that the ``$AdS_3$ wall'' is not consistent with the constraint we need to impose in order to have a unitary theory.

The fields $\gammaup^{+\mu}$ belong to the finite-dimensional $(j_\sl,j_\su)=(-\half,\half)$ representation of $\sltwo\times\sutwo$.
Separating out the contribution~\eqref{Phisu} of $\sutwo$ to $\lambda^{+\mu}$, which has conformal dimension $h=\frac34$, the $\sltwo$ contribution has dimension $h=-\frac{j_\sl(j_\sl-1)}{\nfive}=-\frac34$.  
The $\sltwo$ contribution to $\lambda^{+\mu}$, having $j_\sl=-1/2$, is a non-unitary, non-normalizable operator.%
\footnote{Note that the $j_\sl=-1/2$ representation is rather special; its null vectors provide crucial constraints on the correlation functions that enable their solution by bootstrap methods (see for instance~\rcite{Teschner:1997ft,Giveon:2001up}). }

The chiral components in the mass term~\eqref{ppbar term} are the constraints~\eqref{Qcon} on the fermions $p^{\alpha\beta},\bar p^{\alpha\beta}$.  Since the leading term in the expression is growing without bound in the asymptotic region (because the spin $j_\sl=-1/2$ representation corresponds to a non-normalizable operator), finite action physical states that are predominantly supported in the asymptotic region must satisfy the constraints $\cC^\mu=0$, $\bar \cC^\nu=0$.

For $\nfive>1$, physical states can respond to the UV wall of~\rcite{Seiberg:1999xz} by avoiding it~-- the wavefunctions deform such that they avoid the asymptotic boundary.
But for $\nfive=1$, the wavefunctions are rigid~-- all the states are associated to the $j_\sl=1/2$ representation, which is uniformly spread in $\varphi_\sl$.  They cannot avoid having the dominant part of their support in the region where the fermion mass is diverging.
The mass term~\eqref{massterm} then decouples two linear combinations of the four fermionic pairs $(p^{\alpha\beta},\theta^{\alpha\beta})$ and their antiholomorphic counterparts.%
\footnote{In this respect, it bears some similarity to the R-R plane wave backgrounds studied in~\rcite{Maldacena:2002fy}, 
where the effective dynamics in light-cone gauge yields a massive worldsheet field theory.}
This fact provides a rationale for the constraint as coming directly from the $\PSU$ WZW model~-- the constraint is a free-field approximation to a somewhat richer story.

In the explicit group parametrization used here, $e^{\varphi_\sl}g_\su$ in the first term of~\eqref{ppbar term} combine to make $\lambda^{+\mu}\bar\lambda^{+\nu}$.  Thus, the leading term at large $\varphi_\sl$ in~\eqref{ppbar term} is simply $\sfJ_\con\bar\sfJ_\con$.  We can regard this as a mass term for the linear combination of fermions that doesn't satisfy the constraint, that grows toward the asymptotic boundary.  
It has been established in the $\nfive=1$ string that the path integral for correlators localizes near the asymptotic $AdS_3$ boundary~\cite{Eberhardt:2018ouy, Eberhardt:2019ywk, Eberhardt:2020akk, Eberhardt:2020bgq, Dei:2020zui, Knighton:2020kuh, Dei:2023ivl, Eberhardt:2025sbi} and therefore the correlators will be saturated by field configurations that satisfy the free field constraints, and for all practical purposes the worldsheet fields can be treated as being holomorphic. 

This result explains why the constraint~\eqref{Qcon} and the related mass term~\eqref{ppbar term} don't break any of the spacetime $\cN=(4,4)$ superconformal symmetry, even though they appear to destroy half of the supersymmetry currents $S^{\alpha\beta+}$.  They do not, because the deformation is in fact a modulus which preserves all of this symmetry.  Instead, these supersymmetries are still present, but become composite operators in the worldsheet theory, as in eq.~\eqref{keq1}.

\bigskip
\section{Null gauging}
\label{sec:nullgauge}

Our analysis has brought the realization of $\nfive=1$ $AdS_3\times\bS^3\times\bT^4$ string theory (sourced by a single NS5-brane bound to $n_1$ fundamental strings in the background) into alignment with the standard worldsheet string theory framework. In the process, we have clarified the nature of the theory, and the relation of the free field realization with the $PSU(1,1|2)$ WZW model. 

The formalism we have developed admits a number of further interesting applications.
There are also decoupled fivebrane backgrounds without a large background string charge, that are thus not asymptotically $AdS_3$.  Starting with $\nfive>1$ NS5-branes in asymptotically flat spacetime with string coupling $\gstr$, one scales $\gstr\to0$ while at the same time scaling the fivebrane separations to zero commensurately with $\gstr$.  The resulting fivebrane throat has a linear dilaton, which effectively caps off at the scale of the fivebrane separation, due to the rigidity of string wavefunctions in the throats of isolated fivebranes.  The resulting decoupled dynamics is known as {\it double-scaled little string theory}~\rcite{Giveon:1999px,Giveon:1999tq}.  

A particular $\bZ_\nfive$ symmetric arrangement of the separated branes admits an exactly solvable worldsheet theory.  The initial worldsheet description~\rcite{Giveon:1999px,Giveon:1999tq} of the space transverse to the fivebranes used the coset orbifold $\big(\frac{\SLtwo}{\Uone}\times\frac{\SUtwo}{\Uone}\big)/\bZ_\nfive$.  Subsequently \rcite{Israel:2004ir,Martinec:2017ztd}, it was realized that an equivalent presentation is given by the gauged Wess-Zumino-Witten (WZW) model for
\be
\label{GmodH-2}
\frac\cG\cH = \frac{\SLtwo\times\SUtwo}{\Uone\times\Uone} ~,
\ee
where the gauge group $\cH$ is generated by a pair of null isometries of $\cG$, which we will take to be those generated by the currents
\be
\label{cJdef}
\cJ = J^3+K^3
~~,~~~~
\bar\cJ = \bar J^3+\bar K^3  ~.
\ee
The spectrum of the coset orbifold of~\rcite{Giveon:1999px,Giveon:1999tq}
matches that of the null-gauged WZW model~\eqref{GmodH-2}~\rcite{Israel:2004ir,Giveon:2015raa,Martinec:2017ztd,Dei:2024uyx}.  The null gauging BRST constraint removes a null direction conjugate to that generated by $\cJ$, while motion along $\cJ$ is gauge-redundant.  

Thus, on the worldsheet we start with a (10+2)-dimensional target space%
\footnote{
Here we are allowed to decompactify the longitudinal space of the fivebranes, whereas in the presence of a bunch of fundamental strings that together with the fivebranes generate $AdS_3$, one needs to compactify the fivebrane worldvolume in order to avoid strong coupling issues, due to a factor of $V_4$ in the attractor value of the string coupling in $AdS_3$.}
\begin{equation}
\label{Gtot}
\SLtwo_{n_5} \times \SUtwo_{n_5} \times \bR^{5,1} ~,     
\end{equation}
and gauging removes two directions to get us down to the 9+1 dimensions of the decoupled fivebrane throat.  The choice~\eqref{cJdef} for the null currents removes the two directions from $\SLtwo\times\SUtwo$; the remaining four (spacelike) dimensions describe the transverse space of the fivebranes.

In $AdS_3/CFT_2$ duality, heuristically the time and azimuthal coordinates in $AdS_3$ as well as those of $\bT^4$ parametrize directions along the fivebranes, with the remainder of $AdS_3\times\bS^3$ being transverse.  In the little string theory application, two of the directions in $AdS_3\times\bS^3$ are gauge artifacts, not part of the physical spacetime; and the flat directions along the fivebrane are augmented to $\bR^{5,1}$ (or a compactification thereof). 

Again, in the RNS formalism the worldsheet fermions supply a level $k_\fer=2$ for $\sutwo$, so that it appears that the worldsheet describes only backgrounds with more than one NS5-brane.
We begin by reviewing this construction for $\nfive>1$ in the RNS formalism, and then transcribe the results to the hybrid formalism in preparation for the $\nfive=1$ version in the last subsection.

\subsection[Review of null gauging for \texorpdfstring{$\nfive>1$}{}, in RNS formalism]{Review of null gauging for \texorpdfstring{$\boldsymbol{\nfive>1}$}{}, in RNS formalism}
\label{sec:nullRNS}

The gauging of the null current is implemented by the BRST charge
\be
\label{Qnuldef}
Q_\nul = \oint\!dz\big(\cnul \cJ - \gammanul \Lambda \big) ~, 
\ee
where $\Lambda$ is the superpartner of the null current~\eqref{cJdef} 
\be
\label{Lamdef}
\Lambda = \psi^3 + \chi^3  ~,
\ee
whose role includes removing half the spinor polarizations in 10+2 dimensions in the Ramond sector, in order to have the correct number in 9+1 dimensions.
We have also introduced the null-gauging ghosts
\begin{equation}
\gammanul(z) \betanul(w) \sim \frac{1}{z-w} 
~,\qquad
\cnul(z) \bnul(w) \sim \frac{1}{z-w} ~; 
\end{equation}
$(\bnul,\cnul)$ have conformal dimension $(1,0)$ while $(\betanul,\gammanul)$ have dimension $(\half,\half)$.
These ghosts have stress tensor and supercurrent
\begin{equation}
T^{\gh}_n = \partial \cnul \, \bnul + \frac{1}{2} \gammanul \, \partial \betanul -\frac{1}{2} \partial \gammanul \,  \betanul ~, \qquad
G^{\gh}_n =  -\frac{1}{2} \betanul \partial \cnul + \frac{1}{2}\bnul \gammanul ~, 
\end{equation}
which realize an $\mathcal N=1$ superconformal algebra~\eqref{N=1-superconformal} with central charge ${\mathtt c}^{\gh}_n = -3$. 

We also have the super-Virasoro BRST~\eqref{Qstrdef}; the two are compatible provided the total BRST charge 
\be
Q_\tot= \widehat Q_\str+ Q_\nul
\ee
is nilpotent,
\be
\label{null nilp}
Q_\tot^2= \widehat Q_\str^2+Q_\nul^2+\big\{\widehat Q_\str, Q_\nul\big\} =0 ~,
\ee
where
\begin{align}
\label{12d RNS Qbrst}
\widehat Q_{\str} &= \oint\!dz \Big[ \cstr\big(\widehat T_\m + \hf T^{\gh}_s  + T^{\gh}_n\big) 
- \gammastr\big(\widehat G_\m + \hf G^{\gh}_s + G^{\gh}_n\big) \Big]
\end{align}
and $Q_\nul$ is given in~\eqref{Qnuldef}.  We have decorated the string BRST charge with a hat to denote the fact it has been lifted to a 12d target space, in anticipation of being combined with the null gauge constraint. Similarly, $\widehat T_\m$ and $\widehat G_\m$ denote respectively the matter stress tensor and super stress tensor of the 12d theory, 
\begin{equation}
    \widehat T_\m = T_{\sl} + T_{\su} + T_{\flat} \,,  \qquad    \widehat G_\m = G_{\sl} + G_{\su} + G_{\flat} ~. 
\end{equation}
In a slight abuse of notation, we adopted for the $\bR^{5,1}$ stress tensor $T_{\flat}$ and super stress tensor $G_{\flat}$ the same notation that was used for $\bT^4$. We hope this will not confuse the reader. The expressions \eqref{T4TG}, \eqref{Neq2 G} and \eqref{Neq2 J} are still valid for $\bR^{5,1}$, with the understanding that now $\sfi, \sfj \in \{0,1,2 \}$, instead of just $\sfi, \sfj \in \{1,2 \}$. What has been added (two free bosons, two free fermions and the null-gauging ghosts) has zero net central charge and is superconformal, so does not affect the nilpotence of $\widehat Q_\str$ itself.

The vanishing of the first two terms on the RHS of~\eqref{null nilp} are manifest, resulting from the superconformal symmetry and the (super)gauge symmetry; the vanishing of the last term results from the superconformal invariance of $Q_\nul$ (because~\eqref{Qnuldef} can be written as the superspace integral of a dimension 1/2 superconformal primary).  The physical state space is then the intersection of the cohomologies of $\widehat Q_\str$ and $Q_\nul$.

The null-gauged WZW model thus defined has been studied in~\rcite{Martinec:2017ztd,Martinec:2018nco,Martinec:2019wzw,Martinec:2020gkv,Bufalini:2021ndn,Martinec:2022okx,Bufalini:2022wyp,Bufalini:2022wzu,Dei:2024uyx}.  It provides both an alternative formulation of the little string theory backgrounds of~\rcite{Giveon:1999px,Giveon:1999tq}, as well as a description of particular two-charge BPS backgrounds of decoupled fivebranes.  

We next describe the map of this construction to hybrid variables, and then extend it to $\nfive=1$ using the results of Section~\ref{sec:levelone}.

\subsection[Null gauging for \texorpdfstring{$\nfive>1$}{}, in hybrid formalism]{Null gauging for \texorpdfstring{$\boldsymbol{\nfive>1}$}{}, in hybrid formalism}
\label{sec:nullhybrid}

With the increase in the worldsheet field content to include two extra flat dimensions
\be
\label{extra X}
X^{~}_0, X^*_0 = \frac{1}{\sqrt 2}\big(t\pm y\big)
~,~~~~
\zeta^{~}_0,\zeta^*_0 = \frac{i}{\sqrt 2}\big(\lambda^t\pm\lambda^y\big)
~, 
\ee
as well as the null gauging ghosts, the arena of the RNS\,$\leftrightarrow$\,hybrid map~-- the charge lattice of bosonized spinor fields~-- is enlarged from rank 7 to rank 10.
The additional three directions arise when we bosonize the fields $\lambda^t,\lambda^y$ via 
\begin{equation}
\partial H_6 = \lambda^y \lambda^t 
~,~~~~
\sqrt{2} \, e^{H_6} = \lambda^t+\lambda^y
~,~~~~
\sqrt{2} \, e^{-H_6} =  \lambda^t-\lambda^y  ~, 
\end{equation}
as well as the additional worldsheet spinor ghosts via
\begin{equation}
\label{nulbosbos}
\betanul = e^{-\phinul+\chinul}  \partial \chinul \,, \qquad \gammanul = e^{\phinul-\chinul}~. 
\end{equation} 
We can then promote the RNS $\leftrightarrow$ hybrid map~eq.~\eqref{hybrid map} involving the extra worldsheet spinors to 
\be
\label{hybrid-map-12d}
\begin{aligned}
p^{\alpha \beta} &= e^{\frac{\alpha}{2} H_1 + \frac{\beta}{2} H_2 - \frac{\alpha \beta}{2} H_3 - \frac{1}{2}H_4 - \frac{1}{2}H_5 - \frac{1}{2}H_6 -\frac{1}{2}\phistr + \frac{1}{2} \phinul } ~, \\
\theta^{\alpha \beta} &= e^{\frac{\alpha}{2} H_1 + \frac{\beta}{2} H_2 + \frac{\alpha \beta}{2} H_3 + \frac{1}{2}H_4 + \frac{1}{2}H_5 + \frac{1}{2}H_6 + \frac{1}{2}\phistr - \frac{1}{2} \phinul}  ~, \\
\Psi_1 &= e^{+H_4 + \phistr - \chistr}  ~,\qquad
\Psi^*_1 = e^{-H_4 - \phistr + \chistr}  ~, \\
\Psi_2 &= e^{+H_5 + \phistr - \chistr}  ~,\qquad
\Psi^*_2 = e^{-H_5 - \phistr + \chistr}  ~, \\
\Psi_0 &= e^{+H_6 + \phistr - \chistr }  ~,\qquad
\Psi^*_0 = e^{-H_6 - \phistr + \chistr }  ~, \\
\rhob & = -2 \phistr - H_4 - H_5 - H_6 + \chistr + \phinul  ~, \\
\tilde \rhob & = \phistr + \phinul - \chistr  ~, \\
\chinul & = \chinul ~. 
\end{aligned}
\ee
Once again, this augmented hybrid basis is orthonormal and spans the GSO even sublattice of the lattice of charges in the augmented RNS description.
Note that, just as for the spinor string ghosts, the spinor null ghosts have a picture redundancy.  Aspects of the choice of picture, and the corresponding picture changing operation, are discussed in Appendix~\ref{sec:null pics}.

These variables obey the OPE's
\begin{align}
\label{new-hybrid-OPEs}
p^{\alpha \beta}(z) \theta^{\beta \gamma}(w) & \sim \frac{\vareps^{\alpha\gamma} \vareps^{\beta \delta}}{z-w} ~, \nn\\
\Psi_\sfi(z) \Psi^*_\sfj(w) & \sim \frac{\delta_{\sfi \sfj}}{z-w} ~,  \qquad \sfi, \sfj \in \{ 0, 1, 2\}  ~, \nn\\[.1cm]
\rhob(z) \rhob(w) & \sim - \log(z-w) ~, \\[.1cm]
\tilde \rhob(z) \tilde \rhob(w) & \sim - \log(z-w) ~, \nn\\[.1cm]
\chinul(z) \chinul(w) & \sim \log(z-w) ~, \nn
\end{align}
and (anti-)commute otherwise. Notice that we introduced new hybrid variables $\Psi_0$, $\Psi^*_0$, $\tilde \rhob$ and $\chinul$.  In particular, $\tilde\rhob$ and $\chinul$ form the bosonization of a modified null-gauging ghost $\gammattil,\betattil$ 
\be
\label{tw null gh}
\gammattil = e^{\tilde\rhob-\chinul} ~, \qquad
\betattil = e^{-\tilde\rhob+\chinul}\partial\chinul
\ee
of dimensions (0,1) (as compared to the dimensions $(\hf,\hf)$ of the $(\gammanul,\betanul)$ system of the RNS formalism). 
In the mapping to hybrid variables, the flat $\bR^{5,1}$ directions are twisted such that the dimension 1/2 RNS fermions $\zeta_\sfj,\zeta^*_\sfj$ become fermionic (0,1) systems $\Psi^{~}_\sfj, \Psi^*_\sfj$, and also the ($\tfrac12 , \tfrac12$) system of spinor ghosts $\gammanul,\betanul$ is twisted to the (0,1) system $\gammattil,\betattil$.

The null gauging ghosts admit an $\cN=2$ superconformal algebra~\rcite{Friedan:1985ge}.  In the same way that the $\cN=2$ algebra of the flat directions $X^{~}_\sfj,X^*_\sfj,\zeta^{~}_\sfj,\zeta^*_\sfj$ gets mapped to the topologically twisted $\cN=2$ algebra~\eqref{twisted flat}, the twisted null ghosts $\bnul,\cnul,\betattil,\gammattil$ form a topologically twisted $\cN=2$ algebra~\eqref{twisted null} of central charge ${\mathtt c}^{\rm gh}_n = -3$ which compensates the additional twisted flat directions $\bR^{1,1}$.  The resulting modification $\widehat Q_\str$ of the hybrid string BRST operator is worked out in Appendix~\ref{sec:null hyb brst}.

Under the field redefinition~\eqref{hybrid-map-12d}, the fermionic BRST null constraint for the Coulomb branch of NS5-branes maps to
\begin{equation}
\gammanul \big(\psi^3 + \chi^3\big)  = \sqrt\nfive \, e^{H_3  + \phinul -\chinul}  
= - \sqrt\nfive \,\gammattil\,e^{\rhob}\,\theta^{++}\theta^{--}  ~. 
\label{psi3+chi3}
\end{equation}
Note that while the BRST string operator (as well as the bosonic null current) can be written purely in terms of $\psu$ currents, \emph{without} any need to rely on any particular free field realization, the fermionic null constraint cannot be written in terms of $\psu$ currents only.  Its form depends on the specific (asymptotically) free field realization at hand.  In particular, one of the two $\theta$'s in~\eqref{psi3+chi3} (namely $\theta^{++}$) is absent in the minimal realization~\eqref{keq1} of $\psu_1$, hence the fermionic null constraint can't even be formulated in this case.

This fact provided the initial motivation for our work.  Previously, one had the (asymptotically) free field realization \eqref{Sgens} of $\psu$ at $n_5 >1$ and the minimal free field realization \eqref{keq1} at $n_5=1$, but the relation between them was not understood.  Under this state of affairs, simply expressing the RNS null BRST charge in terms of hybrid variables at $n_5>1$ is not useful since 1)~one does not know how to extend the $n_5>1$ realization of $\psu$ in eq.~\eqref{Sgens} to $n_5 =1$, and 2)~one does not know how to rewrite the $n_5=1$ minimal free field realization \eqref{keq1} in terms of RNS variables.  This logic led to the labyrinthine chain of developments of Section~\ref{sec:levelone} belaboring the various free field realizations at $n_5 =1$.  Without the development of the constrained non-minimal realization of $\psu_1$ in Section~\ref{sec:nonmin}, we would have no way to formulate its null gauging, which provides the worldsheet description of the transverse space to a single fivebrane.  With the non-minimal realization and its RNS\,$\leftrightarrow$\,hybrid map, we can take the RNS formulation of null gauging, process it through the map, and arrive at a hybrid formulation of null gauging.

With these ingredients, we can construct the hybrid formulation of the BRST operator of the null gauged theory.  The usual string BRST operator has the same form~\eqref{J hybrid}, extended to include an extra $\bR^{1,1}$ in the flat directions, as well as the twisted null ghosts; for details, see Appendix~\ref{sec:null details}.

This BRST charge must be combined with the null gauging BRST operator written in hybrid variables, which after the appropriate similarity transformation~\eqref{null similarity} analogous to~\eqref{simtransf} gives
\begin{align}
\widehat Q_\nul &
= \oint\! dz \Big(\cnul\big(J^3+K^3\big) + \sqrt{n_5} \big( \gammattil + \partial \cnul \cstr \big)\, e^{\rhob}\theta^{++}\theta^{--} \Big)(z) ~;
\label{Qnullh}
\end{align}
again, for details see Appendix~\ref{sec:null details}.
Making use of the free field realization in eqs.~\eqref{JK-n5bigger1} and \eqref{Sgens}, it is possible to check directly in terms of hybrid variables that
\begin{equation}
    \{\widehat Q_{\str}, \widehat{Q}_\nul\}=0 ~, 
\end{equation}
as one should expect.

\subsection[Null gauging for \texorpdfstring{$\nfive=1$}{}]{Null gauging for \texorpdfstring{$\boldsymbol{\nfive=1}$}{}}
\label{sec:nullgauge keq1}

Finally, we come to the grand synthesis~-- combining the string BRST constraints of $\widehat Q_\str$ with the non-minimal $\psu$ constraint $Q_\con$ as well as the null constraint $\widehat Q_\nul$.  We can be brief, as almost all the necessary ingredients have already been laid out. 

Our initial interest in complicating the worldsheet theory by introducing extra fields, beyond the minimal set \eqref{minimal-free-fields} needed to realize $\psu$ level one,  lies in the fact that the extra field content is necessary in order to realize the BRST constraint for gauging $\psu$ down to the coset model $\frac{\psu}{\uone\times\uone}$ that describes the transverse space to a single NS5-brane.
The superpartner of the null current~\eqref{psi3+chi3} involves one of the fermions $\theta^{\alpha\beta}$ that is absent from the minimal realization of $\psu_1$.

We include in the BRST operator the null constraint~\eqref{Qnullh} in the non-minimal realization of $\psu_1$.
The compatibility of $\widehat Q_\str$ and $Q_\con$ was established above; the compatibility of $\widehat Q_\str$ and $\widehat Q_\nul$ was established in the RNS formulation~\rcite{Martinec:2017ztd}; so we only need to check the compatibility of $Q_\con$ and $\widehat Q_\nul$.
Nilpotence requires a further modification of the total BRST operator in the simultaneous presence of null gauging and the special $\nfive=1$ constraint 
\begin{align}
\label{Qcon mod2}
Q_\mix &= -\oint\!dz\, \cnul \,e^{\rhob}\big(\gammaup^{++}\theta^{--} + \gammaup^{+-}\gamma^-\theta^{++}\big)  ~.
\end{align}
Since $Q_\mix$ trivially commutes with the similarity transformation current \eqref{null similarity}, it does not need to be modified.  Note that $Q_\mix$ is bilinear in the ghosts $\cnul$ and $\lambda^{+\mu}$, and so only arises when one both gauges the null current and also imposes the constraint $Q_\con$.  It simultaneously represents the fact that fermionic null gauge transformations are modified by the constraint, and that the constraint is modified by fermionic null gauge transformations.  
These modifications are necessary in order that the total BRST operator  
\be
Q_\tot = \widehat Q_\str + Q_\con + \widehat Q_\nul +Q_\mix
\ee
is nilpotent.  Indeed, one can check that $Q_\mix$ is nilpotent, 
\begin{equation}
    Q_\mix^2 = 0 ~, 
\end{equation}
commutes with $Q_\con$ 
\begin{equation}
     \{ Q_\con , Q_\mix \} = 0  ~, 
\end{equation}
and that its failure to commute with the $\gammattil\bnul$ term in $\widehat Q_\str$ is precisely what is needed to cancel the non-commutativity of $Q_\con$ with $\widehat Q_\nul$, 
\begin{equation}
     \quad \{ \widehat Q_\str, Q_\mix \} + \{ Q_\con, \widehat Q_\nul \} = 0 ~. 
\end{equation}
The nilpotence of $Q_\tot$ guarantees that all of the constraints are mutually compatible~-- $\widehat Q_\nul$ reduces the theory to 9+1 dimensions; $Q_\con$ eliminates spurious degrees of freedom in $\psu$; and $\widehat Q_\str$ restricts to the set of physical, on-shell string polarizations.

\bigskip
\section{Interpolating between the fivebrane throat and \texorpdfstring{$ \boldsymbol{AdS_3}$}{}}
\label{sec:supertube}

For $\nfive>1$, the null-gauged WZW model~\eqref{GmodH} describes NS5-branes slightly separated in their transverse space, and suitably modified for $\nfive=1$ it describes the throat of a single NS5-brane.  More generally, one can begin with the full current algebra conformal field theory on
\be
\label{Gtot2}
\cG_\tot = \SLtwo\times\SUtwo\times \bR_t\times\bS^1_y\times\bT^4 ~,
\ee
whose target space is 10+2 dimensional, and look for consistent ways to gauge a null $\cH=U(1)_L\times U(1)_R$ isometry to get down to a 9+1 dimensional target space.  The choice to embed $\cH$ entirely within $\SLtwo\times\SUtwo$ is but one of several possibilities~\rcite{Martinec:2017ztd}.

For $\nfive>1$, a diagonal embedding into $(\sltwo\tight\times\sutwo)\times(\bR_t\tight\times \bS^1_y)$
\be
\label{ST null}
\begin{aligned}
\cnul\cJ &= \cnul\big[(J^3\tight+K^3) - \tfrac{1}{\sqrt 2 } \sfk R_y(i \partial t \tight-i \partial y)\big] ~, \\[.2cm]
\cnulb\bar\cJ &= \cnulb \big[(\bar J^3\tight+\bar K^3) - \tfrac{1}{\sqrt 2 } \sfk R_y(i \bar \partial t \tight + i \bar \partial y)\big] ~, \\[.2cm]
\gammanul\Lambda &=  \gammanul\big[(\psi^3+\chi^3) - \tfrac{1}{\sqrt 2} \sfk R_y(\lambda^t-\lambda^y)\big] 
~, \\[.2cm]
\gammanulb \bar\Lambda &= 
\gammanulb \big[(\bar\psi^3+\bar\chi^3) - \tfrac{1}{\sqrt 2 } \sfk R_y(\bar\lambda^t+\bar\lambda^y)\big] ~, 
\end{aligned}
\ee
leads to a family of backgrounds which interpolate between the linear dilaton throat of the fivebranes in the UV and $AdS_3$ in the IR.%
\footnote{There is in fact a further four-parameter generalization~\rcite{Martinec:2017ztd,Bufalini:2021ndn}, with two additional discrete parameters $\sfm,\sfn$ characterizing additional background rotations along $\bS^3$, but we will for simplicity restrict to the basic family parametrized by $\sfk$ and $R_y$ here.}
The parameter $R_y$ specifies the periodicity $y\sim y\tight+2\pi R_y$ of $\bS^1_y$.
Using the NSR\,$\leftrightarrow$\,hybrid map as in the previous section, one can transcribe into hybrid variables the null currents \eqref{ST null}, 
\be
\label{JST}
\begin{aligned}
\cnul\cJ &= \cnul\big[(J^3\tight+K^3) -  i \sfk R_y \, \partial X_0^* \big] ~, \\[.2cm]
\cnulb\bar\cJ &= \cnulb \big[(\bar J^3\tight+\bar K^3) - i \sfk R_y  \, \bar \partial X_0 \big] ~, \\[.2cm]
\gammanul\Lambda &=  -\gammattil\big(\sqrt\nfive \, e^{\rhob} \theta^{++}\theta^{--} - i \sfk R_y \Psi^*_0 \big) ~,\\[.2cm]
\gammanulb \bar\Lambda &= -\gammattilb\big(\sqrt\nfive \, e^{\bar\rhob} \bar \theta^{++}\bar\theta^{--} - i \sfk R_y \bar\Psi_0 \big) ~. 
\end{aligned}
\ee
Proceeding as in the previous sections and performing the similarity transformation, we finally obtain the null gauging BRST charges 
\be
\label{Qnull final}
\begin{aligned}
   \widehat Q_\nul &= \oint\! dz \Big[\cnul\big(J^3+K^3 -  i \sfk R_y \, \partial X_0^* \big) + \gammattil\big(\sqrt\nfive \, e^{\rhob} \theta^{++}\theta^{--} - i \sfk R_y \Psi^*_0 \big) \\
   & \hspace{150pt} +  \sqrt{n_5}\,  \partial \cnul \cstr \, e^{\rhob}\theta^{++}\theta^{--} - i \sfk R_y \, \partial \cnul \cstr \Psi_0^* \Big] ~, \\
   \widehat{\overline Q}_\nul &=  \oint\! d \bar z \Big[ \cnulb \big(\bar J^3+\bar K^3  - i \sfk R_y  \, \bar \partial X_0 \big) + \gammattilb\big(\sqrt\nfive \, e^{\bar\rhob} \bar \theta^{++}\bar\theta^{--} - i \sfk R_y \bar\Psi_0 \big) \\
   & \hspace{150pt} + \sqrt{n_5} \,  \bar \partial \cnulb \cstrb \, e^{\bar \rhob}\bar \theta^{++} \bar \theta^{--} - i \sfk R_y \, \bar \partial \cnulb \cstrb \bar\Psi_0 \Big] ~. 
\end{aligned}
\ee
Note that to describe a background whose IR is the $AdS_3\times \bS^3$ sourced by NS5-branes and fundamental strings (as opposed to the T-dual background of NS5-branes and momentum), requires a sign flip (axial gauging) in the left and right contributions to the null currents from $\bS^1_y$.  This then requires the opposite of the usual topological twist on the right, so that $\bar\Psi_0$ has conformal dimension one.%
\footnote{Note that due to an unfortunate clash of notations in the literature, the orbifold parameter $\sfk$ in the null gauging coefficients must be distinguished from the level $k=\nfive$ of the WZW model.}

For $\nfive >1$, the resulting background is the fivebrane decoupling limit of the round NS5-F1 supertube of~\rcite{Lunin:2001fv}, as shown in~\rcite{Martinec:2017ztd}.  
The spacetime geometry interpolates between the linear dilaton throat of $\nfive$ fivebranes in the large radius UV region (where the effect of the background F1 charge is a subleading correction), and an  $(AdS_3\times\bS^3)/\bZ_\sfk $ geometry sourced by NS5-F1 flux in the small radius IR region.  The parameter $R_y$ controls the radial position at which the geometry crosses over between the two regimes (in particular, the $AdS_3$ decoupling limit of $AdS_3/CFT_2$ duality takes $R_y\to\infty$); the discrete parameter $\sfk \in \bN$ determines the order of the orbifold twist (if any) of the IR geometry.

Note that in the limit $R_y\to\infty$, the currents being gauged (after an overall rescaling by a factor $1/R_y$) lie entirely along $\bR_t\times\bS^1_y$.  Removing these two directions from $\cG_\tot$ takes us back to string theory on $\SLtwo\times\SUtwo\times\bT^4$.  On the other hand, the limit $R_y\to 0$ removes $\bR_t\times\bS^1_y$ from the gauge currents, and returns us to the construction reviewed in the previous section of the linear dilaton throat of multiple fivebranes.  In the gauge/gravity dual, $1/R_y$ parametrizes a deformation of the spacetime CFT in the direction of little string theory.

In between these two extremes, the role of null gauging in the worldsheet theory is to relate left/right moving dynamics along $\bR_t\times\bS^1_y$, generated by $i(\partial t \tight- \partial y)$, $i(\bar\partial t \tight+ \bar\partial y)$, to that generated by $J^3\tight+K^3,\bar J^3\tight+\bar K^3$ in $\SLtwo\times\SUtwo$.  Of these four directions, there is only one physical time and one physical spatial coordinate; the other two are gauge artifacts.  However, those physical coordinates lie mostly along the geometrical $\bR_t\times\bS^1_y$ cylinder in the UV and mostly along the $\SLtwo\times\SUtwo$ geometry in the IR, and so it is useful to have both coordinates available in the worldsheet description, as \eg\ they directly relate to time and energy as measured by UV observables versus time and energy as measured by IR observables.

Due to the existence of an RNS\,$\leftrightarrow$\,hybrid map for the non-minimal realization of $\psu_1$, there is now no obstruction to a repetition of the same construction for level $\nfive=1$; for $\widehat Q_\nul$ to be nilpotent, any null current will do, and we can in particular use the same choice~\eqref{ST null} that was made for $\nfive>1$. No further modification of $Q_\con$ or $Q_\mix$ is needed.  The theory again has the continuous parameter $R_y$ that determines the scale at which the geometry crosses over from the $(AdS_3\times\bS^3)/\bZ_\sfk $ IR to the linear dilaton UV; the effect of the $\bZ_\sfk $ orbifold on the $\psu_1$ string and its spacetime CFT dual was discussed recently in~\rcite{Gaberdiel:2023dxt}.

The difference with $\nfive>1$ is that now there is a direct relation between the bulk string theory and its gauge theory dual.  For $\nfive=1$, the theory for $R_y\to\infty$ is the bulk string theory dual to the symmetric product orbifold CFT~\rcite{Eberhardt:2018ouy,Eberhardt:2019ywk} in the non-minimal version constructed in Section~\ref{sec:levelone}; the theory for $R_y\to0$ is the bulk string theory dual to abelian little string theory that we constructed in the previous section. 
 The family of theories for finite $R_y$ interpolates between the two.  

\subsection[Bulk dual of the \texorpdfstring{$T \bar T$}{} deformation]{Bulk dual of the \texorpdfstring{$\boldsymbol{T}\bar{\boldsymbol T}$}{} deformation}
\label{sec:TTbar}

There is a close relation between (1) the gauging of a tensor product $\cG\times\cH$ by a diagonally embedded subgroup $\cH$, and (2) the marginal deformation of $\cG$ by a current bilinear~\rcite{Giveon:1993ph,Forste:2003km,Martinec:2023zha}.  The worldsheet $\SLtwo$ sigma model (times $\SUtwo\times\bT^4$), marginally deformed by $J^+\bar J^+$, was studied for $\nfive>1$ in~\rcite{Giveon:2017nie,Asrat:2017tzd} and related to null gauging in~\rcite{Giveon:2017myj}.  The resulting family of worldsheet string models was argued to be the bulk dual of a ``single-trace $T\bar T$ deformation'' of the spacetime dual CFT, though this terminology is rather imprecise~-- away from the weak coupling cusp of its moduli space associated to $\nfive=1$, the spacetime CFT is not a symmetric product orbifold, and the descriptor ``single-trace'' has no meaning.

But for $\nfive=1$, the dual spacetime CFT is in fact the symmetric product orbifold $(\bT^4)^N/S_N$, and the description of its single-trace $T\bar T$ deformation as a current-current deformation of the dual worldsheet string theory was studied in~\rcite{Dei:2024sct}.  Transcribed to the language of null gauging, the work of~\rcite{Dei:2024sct} as well as~\rcite{Giveon:2017nie,Asrat:2017tzd} use an embedding $\cH\subset\cG_\tot$ where the null currents involve $J^+,\bar J^+\in\sltwo$; this choice is adapted to the Poincar\'e slicing of~$AdS_3$.  On the other hand, the choice~\eqref{ST null} involves the timelike generators $J^3,\bar J^3\in\sltwo$, and is adapted to the global slicing of~$AdS_3$.%
\footnote{In Lorentz signature, these two slicings are associated to different states~-- the Poincar\'e slicing corresponds to the massless BTZ black hole, while the global slicing corresponds to the vacuum state (for $\sfk=1$; for $\sfk>1$ one has a $\bZ_\sfk$ conical defect state.}

There is an advantage to working with the equivalent description in terms of a null-gauged WZW model when it comes to computing observables, for instance the correlation functions in the holographic dual (evaluated in the genus expansion of the worldsheet theory).  In the marginally deformed worldsheet description, one must solve for the correlators of the $\sltwo$ theory deformed by a current-current interaction; while in the null-gauged presentation, one simply takes known WZW correlation functions and selects the subset that are gauge-invariant.

We will leave the computation of such correlators, and a full investigation of the spectrum, for future work.  It is relatively straightforward, though, to see the emergence of the $T\bar T$ form of the spectrum, following the analysis of~\rcite{Martinec:2018nco}.

In $AdS_3$ string theory with a pure NS-NS flux background, spectral flow in $\sltwo$ generates sectors of perturbative strings which wind the $AdS_3$ azimuthal direction~\rcite{Maldacena:2000hw}.
However, in the pure fivebrane null gauging of Section~\ref{sec:nullgauge}, this spectral flow operation can be absorbed into a large gauge transformation in $\cH$~\rcite{Martinec:2018nco}, and so does not lead to physically distinct string states, in contrast to the ungauged theory.

With the diagonal null gauging of~\eqref{ST null}, the $\sltwo$ spectral flow sectors are no longer related by large gauge transformations, and hence are physically distinct.  The role of the gauge group now is to relate the null Killing coordinates generated by $J^3+K^3,\bar J^3+\bar K^3$ in $\sltwo\times\sutwo$ to the null coordinate generated by $i(\partial t \tight - \partial y)$ and $i(\bar \partial t \tight + \bar \partial y)$.  The role of large gauge transformations is to impose an equivalence relation between winding in $\sltwo$ and winding along $\bS^1_y$, namely one unit of winding along $\bS^1_y$ is equivalent to $\sfk$ units of winding around $\sltwo$~\rcite{Martinec:2018nco} (in the undeformed symmetric orbifold dual, one is describing excitations around the twisted sector state characterized by the symmetric group conjugacy class where all the cycles have length $\sfk$).

We now impose BRST invariance under $\widehat Q_\str$, $Q_\con$, and the $\widehat Q_\nul$ given by~\eqref{ST null}.  The Virasoro zero-mode constraints impose
\begin{align}
0 &= L_0 - \half = -\frac{j_\sl(j_\sl\tight -1)}{\nfive }+\frac{j_\su(j_\su\tight +1)}{\nfive } - m_\sl w_\sl + m_\su w_\su -\frac\nfive4\big( w_\sl^2\tight -w_\su^2\big) - \frac14\big( E^2\tight -P_y^2\big) + h ~,
\nn\\[.2cm]
0 &= \bar L_0 -\half = -\frac{j_\sl(j_\sl\tight -1)}{\nfive }+\frac{j_\su(j_\su\tight +1)}{\nfive } - \bar m_\sl w_\sl + \bar m_\su \bar w_\su -\frac\nfive4\big( w_\sl^2\tight -\bar w_\su^2\big) - \frac14\big( E^2\tight -\bar P_y^2\big) + \bar h ~,
\end{align}
where $h,\bar h$ are the left/right-moving oscillator excitation levels; we parametrize the zero modes of $t,y$ as
\be
E = w_y R_y + \vareps
~,\qquad
P_y = \frac{n_y}{R_y} + w_y R_y
~,\qquad
\bar P_y = \frac{n_y}{R_y} - w_y R_y ~;
\ee
and $w_\sl,w_\su,\bar w_\su$ label spectral flows in $\sltwo$ and $\sutwo$, which shift the zero mode eigenvalues of $J^3$, $\bar J^3$, $K^3$ and $\bar K^3$ to
\begin{align}
\begin{split}
\label{slsuzm}
M_\sl &= m_\sl+\frac{\nfive}2 w_\sl
~,~~~~\,
\bar M_\sl = \bar m_\sl + \frac{\nfive}2 w_\sl ~,
\\[.1cm]
M_\su &= m_\su+\frac{\nfive}2 w_\su
~,~~~~
\bar M_\su = \bar m_\su + \frac{\nfive}2 \bar w_\su ~.
\end{split}
\end{align}
As discussed above, a linear combination of spectral flows in $\SLtwo$, $\SUtwo$, $\bR_t$ and $\Uone_y$ amounts to a large gauge transformation; we use this freedom to choose a gauge that sets $w_\sl=0$~\rcite{Martinec:2018nco}.

At the same time, one has the null gauge constraints
\begin{align}
\begin{split}
\label{null con}
\sfk(\vareps-n_y) &= 2M_\sl + 2M_\su + {\textit osc.~modes} ~,
\\[.2cm]
\sfk(\vareps+n_y) &= 2\bar M_\sl + 2\bar M_\su + {\textit osc.~modes}  ~,
\end{split}
\end{align}
where the eigenvalues of $J^3,\bar J^3,K^3,\bar K^3$ combine the zero mode eigenvalues~\eqref{slsuzm} 
together with contributions of modes of the charged currents $J^\pm,\bar J^\pm,K^\pm,\bar K^\pm$ and their superpartners.  The gauge constraint~\eqref{null con} relates the two notions of energy conjugate to the two notions of time, as discussed above.
Note that the asymptotic notion of energy given by $\vareps$ is that of the Ramond sector, since spacetime fermions are periodic on $\bS^1_y$.

The Virasoro constraints lead to
(absorbing the contributions of $\sutwo$ spectral flow into $h,\bar h$)
\begin{align}
\vareps &= -w_y R_y +\sqrt{(w_y R_y)^2 +2 (h+\bar h)+ p^2} ~,
\nn\\[.2cm]
p &= \frac{n_y}{R_y} =  \frac{\bar h-h}{w_y R_y}  ~.
\end{align}
The limit of large $R_y$ is the $AdS_3$ decoupling limit; indeed, in this limit, $\vareps R_y\approx (h+\bar h)/w_y$ is the energy of a long string in $AdS_3$, whose excitation level $(h+\bar h)$ is fractionated by the winding $w_y$.

This spectrum has the $T\bar T$ form \rcite{Zamolodchikov:2004ce, Smirnov:2016lqw, Cavaglia:2016oda, Jiang:2019epa}, see for instance~\rcite{Cavaglia:2016oda,Aharony:2018bad}, which is simply the spectrum on a long string of winding $w_y$ on a circle of radius $R_y$.%
\footnote{Specifically, to compare to~\rcite{Aharony:2018bad} eq.~(2.19), one has $(\pi\lambda R)^{-1}\tight=w_yR_y$, $2\pi\lambda RE=h\tight+\bar h$, $P=(h\tight-\bar h)/(w_yR_y)$.  Note, however, that the null-gauged WZW model naturally generates states in the Ramond sector of the spacetime CFT, so energies are measured relative to the Ramond sector BPS bound rather than the NS sector vacuum and its Casimir shift $-\mathtt c/12$.}

The worldsheet formalism describes small excitations around a particular state in the spacetime CFT, and so two different worldsheet theories both with the same asymptotic target space geometry may describe perturbation theory around distinct heavy states.  In the present case, the  analysis of~\rcite{Dei:2024sct} naively describes the $T\bar T$ deformation of the massless BTZ geometry, whereas the null-gauged model of this work naively describes the $T\bar T$ deformation of the maximally spinning Ramond ground state (the spectral flow to the Ramond sector of the CFT vacuum).


\vskip 2cm
\section*{Acknowledgements} 

We thank 
Lorenz Eberhardt, 
Matthias Gaberdiel, 
David Kutasov, 
Kiarash Naderi, 
and
Savdeep Sethi
for discussions.
The research of AD is supported by a Reinhard and Mafalda Oehme Postdoctoral Fellowship in the Enrico Fermi Institute at the University of Chicago.
The work of EJM was supported in part by DOE grant DE-SC0009924. 
%


\appendix



\bigskip
\bigskip
\section{RNS formalism}
\label{sec:RNS review}

In this appendix we spell out our conventions in the RNS formalism and we review the construction of the BRST charge within this formalism for fundamental strings on $AdS_3 \times \bS^3 \times \bT^4$. 
The worldsheet superconformal field theory for type II string theory has an $\mathcal N=1$ superconformal algebra whose components of central charge $\mathtt c$ satisfy the operator product (OPE) algebra
\begin{subequations}
\begin{align}
T(z) T(w) &= \frac{\mathtt c/2}{(z-w)^4} + \frac{2 T(w)}{(z-w)^2} + \frac{\partial T(w)}{z-w} ~, \\
T(z) G(w) &= \frac32\frac{G(w)}{(z-w)^2} + \frac{\partial G}{z-w} ~, \\
G(z) G(w) &= \frac{\mathtt c/6}{(z-w)^3} + \frac{1}{2}\frac{T(w)}{(z-w)} ~.  
\end{align}
\label{N=1-superconformal}
\end{subequations}

\paragraph{Ghosts:} 
Gauging the $\mathcal N=1$ superconformal algebra on the worldsheet leads to a BRST formalism with Grassmann-odd ghosts $(\bstr,\cstr)$ of dimensions $(2,-1)$, and central charge $-26$, and Grassmann-even ghosts $(\betastr, \gammastr)$ with dimensions $(\frac{3}{2},-\frac{1}{2})$ and central charge $11$. The OPE's of the ghost fields read
\begin{equation}
\gammastr(z) \betastr(w) \sim \frac{1}{z-w} 
~,\qquad
\cstr(z) \bstr(w) \sim \frac{1}{z-w} ~. 
\end{equation}
The stress tensor and super stress tensor
\begin{align}
T^{\gh}_s &= T_{\betastr \gammastr} + T_{\bstr \cstr} = \cstr \, \partial \bstr + 2 \partial \cstr \, \bstr - \frac{1}{2} \gammastr \, \partial \betastr -\frac{3}{2} \partial \gammastr \,  \betastr 
~,\qquad \label{T-ghost-RNS} \\
G^{\gh}_s & = -\partial \betastr \cstr -\frac{3}{2} \betastr \partial \cstr + \frac{1}{2}\bstr \gammastr ~, \label{G-ghost-RNS}
\end{align}
realize an $\mathcal N=1$ superconformal algebra~\eqref{N=1-superconformal} with central charge $\mathtt c^{\gh}_s = -15$. The superconformal ghosts $\gammastr$ and $\betastr$ can be bosonized as in eq.~\eqref{ghbos}, while for $\bstr$ and $\cstr$ one simply has
\begin{equation}
    \bstr = e^{-i \sigmab} ~, \qquad \cstr = e^{i \sigmab} ~, \qquad \partial \sigmab = i (\bstr \cstr) ~,
    \label{bscs-bosonization}
\end{equation}
with 
\begin{equation}
    \sigmab(z) \sigmab(w) \sim -\log(z-w) ~. 
\end{equation}

\paragraph{$\boldsymbol{SU(2)_k}$:} 
For the $\mathcal N=1$ $\mathfrak{su}(2)_\nfive$ algebra one has a bosonic current algebra at level $k_\su=\nfive-2$ and a triplet of free fermion superpartners obeying 
\begin{align}
\label{su2 alg}
K^+_\bos(z) K^-_\bos(w) & \sim \frac{\nfive-2}{(z-w)^2} + \frac{2 K^3_\bos(w)}{z-w}  ~~,
&\chi^+(z) \chi^-(w) & \sim \frac{\nfive}{z-w} ~~, \nn\\
K^3_\bos(z) K^3_\bos(w) & \sim -\frac{\nfive-2}{2(z-w)^2} ~~,
&\chi^3(z) \chi^3(w) & \sim \frac{\nfive}{2 (z-w)} ~~, \\ 
K^3_\bos(z) K^\pm_\bos(w) & \sim \frac{K^\pm_\bos}{z-w} ~~, &&\nn
\end{align}
from which one reads off the Killing metric $h_{ab}$ and structure constants $f_{abc}$ of the Lie algebra
\be
h_{ab} = \left(\begin{matrix}0&~\half & ~0 \\ \half &~ 0 & ~0\\ 0&~0&~1\end{matrix}\right)
~~,~~~~
f^{+-}_{~~~~3}=2
~~,~~~~
f^{3+}_{~~~+} = +1
~~,~~~~
f^{3-}_{~~~-} = -1 ~~.
\ee
The stress tensor and super stress tensor read
\begin{subequations}
\label{TGsu RNS}
\begin{align}
\label{Tsu RNS}
T_{\su} &= \frac{1}{2\nfive}\left(K^+_\bos K^-_\bos + K^-_\bos K^+_\bos + 2 K^3_\bos K^3_\bos - \chi^+ \partial \chi^- -\chi^- \partial \chi^+ - 2 \chi^3 \partial \chi^3 \right) ~, \\
G_{\su}  &= \frac{1}{2\nfive}\left( K^+_\bos \chi^- + K^-_\bos \chi^+ + 2 K^3_\bos \chi^3 + \frac{2}{\nfive} \chi^+ \chi^- \chi^3 \right) ~, 
\label{Gsu RNS}
\end{align}
\end{subequations}
and account for a central charge
\begin{equation}
\mathtt c_{\su} = 3 \left( \frac{\nfive-2}{\nfive} + \frac{1}{2} \right) ~. 
\end{equation}
The stress tensor \eqref{Tsu RNS} can be decomposed into its bosonic and fermionic components as 
\begin{equation}
    T_\su = T_\su^\bos + T_\su^\fer ~, 
\end{equation}
where the bosonic $\mathfrak{su}(2)_{k_{\su}}$ stress tensor at level $k_{\su} = n_5 -2$ reads 
\be 
T^\bos_{\su} = \frac{1}{k_{\su}+2}\Bigl( K^3_\bos K^3_\bos + \frac12 K^+_\bos K^-_\bos + \frac12 K^-_\bos K^+_\bos \Bigr) ~,
\label{su-sugawara-T}
\ee
while 
\be 
T_\su^\fer = -\frac{1}{2\nfive}\left(\chi^+ \partial \chi^- + \chi^- \partial \chi^+ + 2 \chi^3 \partial \chi^3 \right) ~.
\ee
Throughout the main text, we will frequently need the Lie algebra generators in the spin $\tfrac12$ representation. They are given by 
\begin{subequations}
\begin{align}
 (\sigma_-)^{\alpha \beta} &= \bigg( \begin{matrix} 1 & 0 \\ 0 & 0 \ \end{matrix} \bigg) \,, &
(\sigma_3)^{\alpha \beta} &= \biggl( \begin{matrix} 0 & 1 \\ 1 & 0 \\ \end{matrix} \bigg) \,, &
(\sigma_+)^{\alpha \beta} &= \bigg( \begin{matrix} 0 & 0 \\ 0 & -1 \\ \end{matrix} \bigg) ~, \\
(\sigma^-)^\alpha{}_\beta &= \bigg( \begin{matrix} 0 & 0 \\ 2 & 0 \ \end{matrix} \bigg) \,, & 
(\sigma^3)^\alpha{}_\beta &= \bigg( \begin{matrix} -1 & 0 \\ 0 & 1 \ \end{matrix} \bigg) \,, &
(\sigma^+)^\alpha{}_\beta &= \bigg( \begin{matrix} 0 & 2 \\ 0 & 0 \ \end{matrix} \bigg) \,, & \\
(\sigma^-)_{\alpha \beta} &= \bigg( \begin{matrix} 2 & 0 \\ 0 & 0 \ \end{matrix} \bigg) \,, & 
(\sigma^3)_{\alpha \beta} &= \bigg( \begin{matrix} 0 & 1 \\ 1 & 0 \ \end{matrix} \bigg) \,, &
(\sigma^+)_{\alpha \beta} &= \bigg( \begin{matrix} 0 & 0 \\ 0 & -2 \ \end{matrix} \bigg) \,, & \\
(\sigma_-)^\alpha{}_\beta &= \bigg( \begin{matrix} 0 & 1 \\ 0 & 0 \ \end{matrix} \bigg) \,, & 
(\sigma_3)^\alpha{}_\beta &= \bigg( \begin{matrix} -1 & 0 \\ 0 & 1 \ \end{matrix} \bigg) \,, &
(\sigma_+)^\alpha{}_\beta &= \bigg( \begin{matrix} 0 & 0 \\ 1 & 0 \ \end{matrix} \bigg) \,, & \\
(\sigma_-)_{\alpha \beta} &= \bigg( \begin{matrix} 0 & 0 \\ 0 & -1 \ \end{matrix} \bigg) \,, & 
(\sigma_3)_{\alpha \beta} &= \bigg( \begin{matrix} 0 & 1 \\ 1 & 0 \ \end{matrix} \bigg) \,, &
(\sigma_+)_{\alpha \beta} &= \bigg( \begin{matrix} 1 & 0 \\ 0 & 0 \ \end{matrix} \bigg) \,,  & 
\end{align}
\label{sigma123}
\end{subequations}
\ie\ spinor indices are raised and lowered with 2d Levi-Civita symbol $\varepsilon^{+-} = - \varepsilon_{+-} =1$.

\paragraph{$\boldsymbol{SL(2,\mathbb R)_\nfive}$:} 
The $AdS_3$ factor in target space is described by the
$\mathcal N=1$ $\mathfrak{sl}(2,\mathbb R)_\nfive$ current algebra 
\begin{align}
\label{sl2 alg}
J^+_\bos(z) J^-_\bos(w) & \sim \frac{\nfive+2}{(z-w)^2} - \frac{2 J^3_\bos(w)}{z-w} ~~,
&\psi^\pm(z)  \psi^\mp(w) & \sim \frac{\nfive}{z-w} ~~, \nn\\
J^3_\bos(z) J^3_\bos(w) & \sim -\frac{\nfive+2}{2(z-w)^2} ~~,
&\psi^3(z) \psi^3(w) & \sim -\frac{\nfive}{2 (z-w)} ~~,\\ 
J^3_\bos(z) J^\pm_\bos(w) & \sim \frac{J^\pm_\bos}{z-w} ~~, && \nn
\end{align}
from which one reads off the Killing metric $h_{ab}$ and structure constants $f_{abc}$ of the Lie algebra
\be
h_{ab} = \left(\begin{matrix}~0&~\half & 0 \\ ~\half &~ 0 &0\\ ~0&~0&-1\end{matrix}\right)
~~,~~~~
f^{+-}_{~~~~3}=-2
~~,~~~~
f^{3+}_{~~~+} = +1
~~,~~~~
f^{3-}_{~~~-} = -1 ~~. 
\ee
The stress tensor and super stress tensor
\begin{subequations}
\label{TGsl RNS}
\begin{align}
T_{\sl} &= \frac{1}{2\nfive}\left(J^+_\bos J^-_\bos + J^-_\bos J^+_\bos - 2 J^3_\bos J^3_\bos - \psi^+ \psi^- -\psi^- \partial \psi^+ + 2 \psi^3 \partial \psi^3 \right) ~, \label{Tsl-RNS}\\
G_{\sl}  &= \frac{1}{2\nfive}\left( J^+_\bos \psi^- + J^-_\bos \psi^+ - 2 J^3_\bos \psi^3 -\frac{2}{\nfive} \psi^+ \psi^- \psi^3 \right) ~, 
\label{Gsl RNS}
\end{align}
\end{subequations}
form a superconformal algebra with central charge 
\begin{equation}
\mathtt c_{\sl} = 3 \left( \frac{\nfive+2}{\nfive} + \frac{1}{2} \right) ~. 
\end{equation}
The stress tensor \eqref{Tsl-RNS} can be decomposed as 
\be
T_\sl = T_\sl^\bos + T_\sl^\fer ~, 
\ee
where the $\mathfrak{sl}(2,\mathbb R)$ bosonic stress tensor at level $k_{\sl}=n_5+2$ reads 
\begin{equation}
T^\bos_{\sl} = \frac{1}{k_\sl-2}\Bigl( -J^3_\bos J^3_\bos + \frac12 J^+_\bos J^-_\bos + \frac12 J^-_\bos J^+_\bos \Bigr) ~, 
\label{sl-sugawara-T}
\end{equation}
and 
\be
T^\fer_{\sl} = -\frac{1}{2\nfive}\left(\psi^+ \psi^- + \psi^- \partial \psi^+ - 2 \psi^3 \partial \psi^3 \right) ~. 
\ee
Finally, the Lie algebra generators in the spin $\tfrac12$ representation are given by 
\begin{subequations}
\label{tau123}
\begin{align}
 (\tau_-)^{\alpha \beta} &= \bigg( \begin{matrix} -1 & 0 \\ 0 & 0 \ \end{matrix} \bigg) \,, &
(\tau_3)^{\alpha \beta} &= \biggl( \begin{matrix} 0 & 1 \\ 1 & 0 \\ \end{matrix} \bigg) \,, &
(\tau_+)^{\alpha \beta} &= \bigg( \begin{matrix} 0 & 0 \\ 0 & -1 \\ \end{matrix} \bigg) \,, &\\
(\tau^-)^\alpha{}_\beta &= \bigg( \begin{matrix} 0 & 0 \\ -2 & 0 \ \end{matrix} \bigg) \,, & 
(\tau^3)^\alpha{}_\beta &= \bigg( \begin{matrix} -1 & 0 \\ 0 & 1 \ \end{matrix} \bigg) \,, &
(\tau^+)^\alpha{}_\beta &= \bigg( \begin{matrix} 0 & 2 \\ 0 & 0 \ \end{matrix} \bigg) \,, & \\
(\tau^-)_{\alpha \beta} &= \bigg( \begin{matrix} -2 & 0 \\ 0 & 0 \ \end{matrix} \bigg) \,, & 
(\tau^3)_{\alpha \beta} &= \bigg( \begin{matrix} 0 & 1 \\ 1 & 0 \ \end{matrix} \bigg) \,, &
(\tau^+)_{\alpha \beta} &= \bigg( \begin{matrix} 0 & 0 \\ 0 & -2 \ \end{matrix} \bigg) \,, & \\
(\tau_-)^\alpha{}_\beta &= \bigg( \begin{matrix} 0 & -1 \\ 0 & 0 \ \end{matrix} \bigg) \,, & 
(\tau_3)^\alpha{}_\beta &= \bigg( \begin{matrix} -1 & 0 \\ 0 & 1 \ \end{matrix} \bigg) \,, &
(\tau_+)^\alpha{}_\beta &= \bigg( \begin{matrix} 0 & 0 \\ 1 & 0 \ \end{matrix} \bigg) \,, & \\
(\tau_-)_{\alpha \beta} &= \bigg( \begin{matrix} 0 & 0 \\ 0 & 1 \ \end{matrix} \bigg) \,, & 
(\tau_3)_{\alpha \beta} &= \bigg( \begin{matrix} 0 & 1 \\ 1 & 0 \ \end{matrix} \bigg) \,, &
(\tau_+)_{\alpha \beta} &= \bigg( \begin{matrix} 1 & 0 \\ 0 & 0 \ \end{matrix} \bigg) \,.  & 
\end{align}
\end{subequations}

\paragraph{Flat space:} 
We consider four free bosons and four free fermions in complex pairs
\begin{equation}
\partial X^{~}_\sfi(z) \partial X^*_{\sfj}(w) \sim \frac{\delta_{\sfi \sfj}}{(z-w)^2}  
~,\qquad
\zeta^{~}_\sfi(z)  \zeta^*_{\sfj}(w) \sim \frac{\delta_{\sfi \sfj}}{z-w} ~, 
\label{T4-RNS-OPEs}
\end{equation}
where $\sfi, \sfj \in \{1,2\}$.    
The stress tensor and supercurrent 
\be
\label{T4TG}
\begin{aligned}
T_{\flat} &= T_{\flat}^\bos + T_{\flat}^\fer =   \partial X^\sfj \partial  X^*_\sfj  - \frac12 (\zeta^\sfj \partial  \zeta^*_\sfj + \zeta^*_\sfj \partial \zeta^\sfj )  ~, \\
G_{\flat} &= \frac{1}{2}\left( \partial X^\sfj  \zeta^*_\sfj + \partial  X^*_\sfj \zeta^\sfj  \right) ~, 
\end{aligned}
\ee
close into the $\mathcal N=1$ superconformal algebra with $\mathtt c_{\flat} = 6$ for $\bT^4$. 

The two terms in the supercurrent~\eqref{T4TG} are separately the fermionic generators of an $\cN=2$ superconformal algebra
\be
\label{Neq2 G}
G^+_{\flat} = \partial X^*_\sfj\zeta^\sfj
~,\qquad
G^-_{\flat} = \partial X^{\vphantom{|}}_\sfj\zeta^*_\sfj  ~,
\ee
together with the stress tensor and $U(1)_\cR$ current
\be\label{Neq2 J}
J_{\flat} = \half \zeta^\sfj\zeta^*_\sfj  ~,
\ee
where in our conventions the $\mathcal N=2$ superconformal algebra reads
\begin{subequations}
\begin{align}
    T (z) \, T (w) &= \frac{\mathtt c /2}{(z-w)^4} + \frac{2 T (w)}{(z-w)^2} + \frac{\partial T (w)}{z-w} ~, \\
    T (z) \, G^\pm (w) &= \frac{3}{2}\frac{G^\pm (w)}{(z-w)^2} + \frac{\partial G^\pm (w)}{z-w} ~, \\
    T (z) \, J (w) &= \frac{J (w)}{(z-w)^2} + \frac{\partial J (w)}{z-w} ~, \\
    J (z) \, G^\pm (w)  &= \pm \frac{1}{2} \frac{G^\pm (w)}{(z-w)} ~, \\
    J (z) \, J (w) & = \frac{\mathtt c /12}{(z-w)^2} ~, \\
    G^+ (z) \, G^- (w) &= \frac{\mathtt c /3}{(z-w)^3} + \frac{2 J (w)}{(z-w)^2} + \frac{T (w) + \partial J (w) }{(z-w)} ~. \label{Gp-Gm-untwisted-OPE}
\end{align}
\end{subequations}

\paragraph{BRST charge:} 
We define the matter stress tensor and supercurrent
\begin{subequations}
\begin{align}
T_\m &= T_{\sl}+T_{\su}+T_{\flat} ~, \label{T-matter-RNS}\\ 
G_\m &= G_{\sl}+G_{\su}+G_{\flat} ~. 
\end{align}
\end{subequations}
The BRST current reads
\begin{equation}
\label{JBRST-10d-RNS}
\sfJ_\brst  = \cstr \left(T_\m +\frac{1}{2}T^{\gh}_s \right) - \gammastr \left(  G_\m +\frac{1}{2}G^{\gh}_s \right) ~, 
\end{equation}
where $T^\gh_s$ and $G^\gh_s$ have been defined in eqs.~\eqref{T-ghost-RNS} and \eqref{G-ghost-RNS} respectively. The BRST current can be decomposed as 
\begin{equation}
\sfJ_\str = \sfJ_0 + \sfJ_1 + \sfJ_2 + \frac{3}{4} \partial(\gammastr \betastr \cstr) ~, 
\label{RNS-BRST-decomposed}
\end{equation}
where 
\begin{align}
\label{Jbrst RNS}
\sfJ_0 & = \cstr (T_\m +T^{\gh}_s + \partial \cstr \, \bstr) ~, \nn\\[.2cm]
\sfJ_1 & = - \gammastr G_\m ~, \\
\sfJ_2 & = -\frac{1}{4} \bstr \gammastr^2 ~. \nn
\end{align}
One can check that the first order pole in the OPE of $\sfJ_\str$ with itself is a total derivative and hence that the BRST charge
\be
\label{Qstrdef}
Q_\str = \oint\! d z \,  \sfJ_\brst (z)
\ee
squares to zero.

\paragraph{The RNS stress tensor:} The total stress tensor with zero central charge in the RNS formalism is simply given by the sum of the matter stress tensor \eqref{T-matter-RNS} and the ghost stress tensor \eqref{T-ghost-RNS},
\be 
T^{\it RNS} = T_\m + T^\gh_s ~. 
\ee

\section{Details on the hybrid formalism and its map to RNS}
\label{sec:hybrid map}

In this appendix, we collect some details of the hybrid formalism, and its equivalence to the RNS formalism reviewed in the previous appendix.  We begin with the latter, rewriting eq.~\eqref{hybrid map} as
\begin{align}
p^{++} &= e^{\sfv_1\cdot\sfH} \,, 
& p^{+-} &= e^{\sfv_2\cdot\sfH} \,, 
& p^{-+} &= e^{\sfv_3\cdot\sfH} \,, 
& p^{--} &= e^{\sfv_4\cdot\sfH} \,, 
\nn\\
\theta^{++} &= e^{-\sfv_1\cdot\sfH} \,, 
& \theta^{+-} &= e^{-\sfv_2\cdot\sfH} \,, 
& \theta^{-+} &= e^{-\sfv_3\cdot\sfH} \,, 
& \theta^{--} &= e^{-\sfv_4\cdot\sfH} \,, 
\\
\Psi_1 &= e^{\sfv_5\cdot\sfH} \,, 
& \Psi^*_1 &= e^{-\sfv_5\cdot\sfH} \,, 
& \Psi_2 &= e^{\sfv_6\cdot\sfH} \,, 
& \Psi^*_2 &= e^{-\sfv_6\cdot\sfH} \,, 
\nn\\
\rhob &= \sfv_7\cdot\sfH \,, &&&&&& \nn
\end{align}
where we introduced the 7-dimensional basis of RNS fields $\sfH=(H_1,H_2,H_3,H_4,H_5,\phistr,\chistr)$ and wrote the hybrid basis in terms of the RNS basis as
\begin{align}
\label{hybrid-basis}
\sfv_1&=\frac{1}{2} (1,1,-1,-1,-1,-1,0) ~~, 
&\sfv_5&=(0,0,0,1,0,1,-1) ~~, \nn\\
\sfv_2&=\frac{1}{2} (1,-1,1,-1,-1,-1,0) ~~, 
&\sfv_6&=(0,0,0,0,1,1,-1) ~~, \\
\sfv_3&=\frac{1}{2} (-1,1,1,-1,-1,-1,0) ~~,
&\sfv_7&=(0,0,0,-1,-1,-2,1) ~~, \nn\\ 
\sfv_4&=\frac{1}{2} (-1,-1,-1,-1,-1,-1,0) ~~. && \nn
\end{align}

In the type II theory in RNS variables, one has a chiral GSO projection, which can be written in terms of the bosonization above as a restriction on the lattice vectors $\sfv$~\rcite{Giveon:1998ns}.  The primitive vectors for spin fields $\sfv=\half(\alpha_1,...,\alpha_6,0)$ with $\alpha_i=\pm1$ satisfy the conditions 
\be
\prod_{i=1}^6 \alpha_i = +1
~~,~~~~
\prod_{i=1}^3 \alpha_i = -1 ~,
\ee
and similarly for right-movers.
The first condition is the GSO projection; the second condition arises from the three-fermion terms in~\eqref{Gsu RNS}, \eqref{Gsl RNS}.

The vectors \eqref{hybrid-basis} span the lattice of GSO-even vertex operators, 
are linearly independent and orthonormal, 
\begin{equation}
g_{ab} = \sfv_a^i \eta_{ij} \sfv_b^j = \text{diag}(1,1,1,1,1,1,-1) ~~, 
\end{equation}
with respect to the inner product of the RNS basis fields 
\begin{equation}
\eta_{ij} = \text{diag}(1,1,1,1,1,-1,1) ~~. 
\end{equation}
The two field bases are thus completely equivalent.

\subsection{Sugawara stress tensors}

In hybrid variables, the theory factorizes into a WZW model for $\PSU$, a ``topologically twisted'' $\bT^4$, a free field $\rhob$, and the reparametrization ghosts $\bstr,\cstr$ (or equivalently $\sigmab$).  
The total RNS stress tensor can be rewritten in terms of the hybrid variables \eqref{hybrid map} as
\begin{align}
\cT^{\text{Hybrid}} & = \cT_{\mathfrak{psu}} + \mathcal T_{\flat} + \mathcal T_\rhob  +  T_{\bstr \cstr} ~,
\end{align}
where
\begin{subequations}
\begin{align}
    \cT_{\mathfrak{psu}} &= T_{\sl}^{\bos} + T_{\su}^{\bos} +\mathcal T_{p \theta} ~, \label{Tpsu}\\[.1cm]
    \mathcal T_{p \theta} & = 
    -\vareps_{\alpha\beta}\vareps_{\mu\nu}\,p^{\alpha\mu}\partial\theta^{\beta\nu}~,\\
    \mathcal T_\Psi &= -  \Psi^*_\sfj \partial \Psi^\sfj ~, \\[.1cm]
    \mathcal T_{\flat} & = T_{\flat}^{\bos}  + \mathcal T_\Psi ~, \\[.1cm]
    \mathcal T_\rhob &= -\tfrac{1}{2} (\partial \rhob)^2 + \tfrac{3}{2}\partial^2 \rhob ~. \label{Trho}
\end{align}
\label{hybrid-stress-tensors}%
\end{subequations}
The $\mathfrak{psu}(1,1|2)_{n_5}$ stress tensor can be expressed directly in terms of the $\psu$ currents as 
\begin{equation}
\mathcal T_{\mathfrak{psu}} = \frac{1}{n_5}\Bigl( -J^3 J^3 + \tfrac12 J^+J^- + \tfrac12 J^- J^+ + K^3 K^3 + \tfrac12 K^+K^- + \tfrac12 K^-K^+ + \vareps_{\alpha \beta} \vareps_{\gamma \delta} \vareps_{\rho \sigma} S^{\alpha\gamma\rho}S^{\beta\delta\sigma} \Bigr) ~. 
\label{psu-sugawara-stress-tensor}
\end{equation}
In fact, one can check that eq.~\eqref{psu-sugawara-stress-tensor} obeys the OPE one expects for a stress tensor of central charge $c_{\mathfrak{psu}} = -2$ and that replacing eqs.~\eqref{JK-n5bigger1} and \eqref{Sgens} into eq.~\eqref{psu-sugawara-stress-tensor} reproduces eq.~\eqref{Tpsu}. 

\subsection[The twisted \texorpdfstring{$\mathcal N=2$}{} algebra on the worldsheet]{The twisted \texorpdfstring{$\boldsymbol{\mathcal N=2}$}{} algebra on the worldsheet}

One can ``topologically twist'' the $\cN=2$ algebra~\eqref{Neq2 G}, \eqref{Neq2 J} in either of two ways, defining twisted stress tensors
\be
\cT_\pm = T\pm  J  ~; \label{twisted Ts}
\ee
one has this choice independently for holomorphic and anti-holomorphic fields, and more generally, separately for each term in the sum over $\sfj$ in~\eqref{Neq2 G}, \eqref{Neq2 J}.
With the twisting $\cT_+$, the fields $(\zeta_\sfj,\zeta^*_\sfj)$ have dimension $(0,1)$, while for the twisting $\cT_-$ they have dimension $(1,0)$. 
With the choice $\mathcal T = \mathcal T_+$, in our conventions the twisted $\mathcal N=2$ algebra reads
\begingroup
\allowdisplaybreaks
\begin{subequations}
\label{N=2-twisted-T+}
\begin{align}
    \mathcal T_+(z) \, \mathcal T_+(w) &=  \frac{2 \, \mathcal T_+(w)}{(z-w)^2} + \frac{\partial \mathcal T_+(w)}{(z-w)} ~, 
    \\
    \mathcal T_+(z) \, \mathcal G^+(w) &= \frac{\mathcal G^+(w)}{(z-w)^2} + \frac{\partial \mathcal G^+(w)}{(z-w)} ~, 
    \label{eqb}\\
    \mathcal T_+(z) \, \mathcal G^-(w) &= \frac{2 \, \mathcal G^-(w)}{(z-w)^2} + \frac{\partial \mathcal G^-(w)}{(z-w)} ~, \\
    \mathcal T_+(z) \, \mathcal J^\mathcal{R}(w) &= -\frac{\mathtt c/6}{(z-w)^3} + \frac{\mathcal J^{\mathcal R}(w)}{(z-w)^2} + \frac{\partial \mathcal J^{\mathcal R}(w)}{(z-w)} ~, \\
    \mathcal J^{\mathcal R}(z) \, \mathcal G^\pm(w)  &= \pm \frac{1}{2} \frac{\mathcal G^\pm(w)}{(z-w)} ~, \\
    \mathcal J^{\mathcal R}(z) \, \mathcal J^{\mathcal R}(w) & = \frac{\mathtt c/12}{(z-w)^2} ~, \\
    \mathcal G^+(z) \, \mathcal G^-(w) &= \frac{\mathtt c/3}{(z-w)^3} + \frac{2 \mathcal J^{\mathcal R}(w)}{(z-w)^2} + \frac{\mathcal T_+(w)}{(z-w)} ~. \label{Gp-Gm-twisted-OPE}
\end{align}
\end{subequations}
\endgroup

If instead one defines the twisted stress tensor as $\cT = \cT_-$, one arrives at an almost identical algebra, but now with $\cG^+$ having conformal dimension two while $\cG^-$ has dimension one; also the leading (central charge) term in the $\cT$-$\cJ$ OPE flips sign, and one replaces $\cT_+= \cT_-+2\partial\cJ^{\cR}$ on the RHS of~\eqref{Gp-Gm-twisted-OPE}.
For the most part, we will make the choice $\mathcal T = \mathcal T_+$ of twist direction, and suppress the subscript on~$\cT$.

\paragraph{The flat space twisted $\boldsymbol{\mathcal N=2}$ algebra:}
The hybrid fields $\Psi^{~}_\sfi,\Psi^*_\sfi$ of~\eqref{hybrid map} and~\eqref{hybrid-map-12d} together with the corresponding bosons $X_\sfi ,X^*_\sfi$ realize precisely the topologically twisted $\cN=2$ algebra \eqref{N=2-twisted-T+} on $\mathbb T^4$. The twisted $\cN=2$ currents and supercurrents can be expressed in terms of the $\mathbb T^4$ twisted free fields as
\be
\label{twisted flat}
\begin{aligned}
\cT_{\flat} &= \partial X^\sfj \partial X^*_\sfj -  \Psi^*_\sfj \partial \Psi^\sfj ~, & \cJ^{\mathcal R}_{\flat} &= \frac12 \Psi^\sfj \Psi^*_\sfj ~, \\
\cG^+_{\flat} &= \partial X^*_\sfj \Psi^\sfj ~, & \cG^-_{\flat} &= \partial X^\sfj \Psi^*_\sfj  ~,
\end{aligned}
\ee
where $\sfj=1,2$ runs over the two complex directions of $\bT^4$.

\vskip .5cm
\paragraph{The full $\boldsymbol{\mathcal{N}=2}$ algebra on the worldsheet:} 

Before the similarity transformation~\eqref{simtransf} the $\mathcal N=2$ algebra on the worldsheet reads
\begin{subequations}
\begin{align}
    \cT &= \cT_{\mathfrak{psu}} + \mathcal T_{\flat} + \mathcal T_\rhob  +  T_{\bstr \cstr} ~, \\
    \cG^+ &= \mfJ_0 + \mfJ_1 + \mfJ_2 ~, \\
    \cG^- &= \bstr ~, \\
    \cJ^{\mathcal R} &= \frac{1}{2} (\partial \rhob + i \partial \sigmab) +  \cJ^{\mathcal R}_{\flat}    ~,  
\end{align}
\end{subequations}
where $\mfJ_0$, $\mfJ_1$ and $\mfJ_2$ are given in eq.~\eqref{hyb-BRST-before-similarity}. The similarity transformation~\eqref{simtransf} does not affect $\cT$ and $\cJ^{\mathcal R}$, modifies $\cG^+$ as described around eq.~\eqref{J hybrid} and adds $- \cG_{\flat}^-$ to the $\cG^-$ supercurrent,
\begin{subequations}
\begin{align}
    \cT &= \cT_{\mathfrak{psu}} + \mathcal T_{\flat} + \mathcal T_\rhob  +  T_{\bstr \cstr} ~, \\
    \cG^+ &= \mfJ_0 + \mfJ_1 + \mfJ_2 ~, \\
    \cG^- &= \bstr - \cG_{\flat}^- ~, \\
    \cJ^{\mathcal R} &= \frac{1}{2} (\partial \rhob + i \partial \sigmab) +  \cJ^{\mathcal R}_{\flat}    ~,  
\end{align}
\end{subequations}
where $\mfJ_0$, $\mfJ_1$ and $\mfJ_2$ are now given in eq.~\eqref{J hybrid} and fields are normal ordered from the right to the left, e.g.\footnote{We thank Kiarash Naderi for explaining this to us.}
\be
\mfJ_0 = (T_{\mathfrak{psu}} \, e^{i\sigmab}) - \hf \Bigl(\partial(\rhob\tight+ i\sigmab) \bigl(\partial(\rhob\tight+ i\sigmab) e^{i\sigmab} \bigr) \Bigr) + \hf \bigl(\partial^2(\rhob\tight+ i \sigmab) e^{i\sigmab} \bigr) ~, 
\ee
and the term $e^{-\rhob} J^a p^{\alpha\mu}p^{\beta\nu}$ entering eq.~\eqref{J1 hybrid} is normal ordered as 
\be 
\Bigl( e^{-\rhob} \bigl( J^a (p^{\alpha\mu}p^{\beta\nu}) \bigr) \Bigr) ~. 
\ee

\subsection{The null vector}
\label{sec:the-null-vector-app}

We now spell out the state $\Sigma$ entering equation \eqref{null-vector-Qcon-exact}, in terms of the non-minimal free fields $\beta^+, \gamma^-, p^{\alpha\beta}, \theta^{\alpha \beta}, \lambda^{+\alpha}, \pi^{-\alpha}$. We obtain 
\begin{align}
\Sigma &= -\beta^+\pi^{--}\theta^{-+} + \beta^+\pi^{-+}\theta^{--} -\gamma^- \beta^+\pi^{--}\theta^{++} + \gamma^-\beta^+\pi^{-+}\theta^{+-} -\tfrac{1}{3} \lambda^{+-}\pi^{-+}\pi^{--}\theta^{++} \nn \\
& \hspace{12pt} + \tfrac{1}{3}\lambda^{+-}\pi^{-+}\pi^{-+}\theta^{+-} + 
 \tfrac{1}{3} \lambda^{++} \pi^{--} \pi^{--} \theta^{++} - 
 \tfrac{1}{3}  \lambda^{++} \pi^{-+} \pi^{--} \theta^{+-} + 
 \tfrac{4}{3}   \pi^{--} p^{+-} \theta^{-+} \theta^{++} \nn \\
 & \hspace{12pt} -  \tfrac{2}{3}   \pi^{--} p^{++} \theta^{--} \theta^{++} - \tfrac{2}{3}   \pi^{--} p^{++} \theta^{-+} \theta^{+-} - 
 \tfrac{2}{3}  \pi^{-+} p^{+-}  \theta^{--}  \theta^{++} - 
 \tfrac{2}{3}  \pi^{-+} p^{+-}  \theta^{-+} \theta^{+-} \nn \\ 
 & \hspace{12pt} +\tfrac{4}{3}  \pi^{-+} p^{++}  \theta^{--} \theta^{+-} + \partial \pi^{--} \theta^{++} - \partial \pi^{-+} \theta^{+-} ~, 
 \label{Sigma}
\end{align}
where fields are normal ordered from the right to the left, \eg~$\beta^+\pi^{--}\theta^{-+} = (\beta^+(\pi^{--}\theta^{-+}))$. 


\section{Details on the representation theory of \texorpdfstring{$\boldsymbol{\psu_1}$}{}}
\label{sec:details-repr-theory}

In this appendix we complement the discussion in Section~\ref{sec:representation-theory} of the main text about $\psu_1$ representations in terms of non-minimal free fields. In particular, in Appendix~\ref{sec:repr-th-coninuous} we show that when both $\lambda_2 \neq \tfrac12$ and $\lambda_3 \neq \tfrac12$, all the states that are $Q_\con$-closed are also $Q_\con$-exact. Similarly, we show in Appendix~\ref{sec:truncated-reps} that the same happens when \eg~$\lambda_2 \neq \tfrac12$ and $\lambda_3 =\tfrac12$. Finally, in Appendix~\ref{sec:doubly-truncated-reps} we show that for $\lambda_2 = \lambda_3 = \tfrac12$ the $Q_\con$ cohomology only contains the states~\eqref{non-min-rep-0-singlet}. 

\subsection{Continuous representations}
\label{sec:repr-th-coninuous}

In this section, we consider the continuous representations \eqref{zero-mode-rep} of the free field zero-mode algebra \eqref{zero-mode-algebra} for $\lambda_2 \neq \tfrac12$ and $\lambda_3 \neq \tfrac12$. We are going to show that none of the states in continuous representations are part of the $Q_\con$ cohomology. We order the discussion according to the number of $\theta^{\alpha \beta}$ modes.

\paragraph{No $\boldsymbol{\theta^{\alpha \beta}} $ modes:} All states of the form $\ket{m_1, m_2, m_3}$ are $Q_\con$-exact since 
\be
\label{zero-thetas-Qcon-exact-1}
Q_\con \, \theta^{++}_0 \ket{m_1, m_2, m_3+1} = - \lambda^{++}_0 \{ p^{--}_0, \theta^{++}_0 \} \ket{m_1, m_2, m_3+1} = - \ket{m_1, m_2, m_3} ~, 
\ee
or equivalently
\begin{align}
Q_\con \, \theta^{+-}_0 \ket{m_1, m_2+1, m_3} = \ket{m_1, m_2, m_3} ~. 
\label{zero-thetas-Qcon-exact-2}
\end{align}

\paragraph{One $\boldsymbol{\theta^{\alpha \beta}}$ mode:} Let us consider linear combinations of states of the form $\theta^{\alpha \beta}_0 \ket{m_1, m_2, m_3}$. More precisely, in order for states to have definite $J^3_0$ and $K^3_0$ charges, we consider the linear combination
\begin{multline}
a_{++} \, \theta^{++}_0 \ket{m_1, m_2, m_3 + 1} + a_{+-} \, \theta^{+-}_0 \ket{m_1, m_2 + 1, m_3}  \\
+ a_{-+} \, \theta^{-+}_0 \ket{m_1 + 1, m_2, m_3 + 1} + a_{--} \, \theta^{--}_0 \ket{m_1 + 1, m_2 + 1, m_3} ~,
\label{one-theta-ansatz}
\end{multline}
where $a_{++}, a_{+-}, a_{-+}, a_{--}$ are so far unrestricted real coefficients. Requiring the linear combination \eqref{one-theta-ansatz} to be $Q_\con$-closed we find 
\be 
\vareps^{\alpha \beta} a_{\alpha \beta} = 0 ~. 
\label{one-theta-coeff}
\ee
Without loss of generality, and renaming $m_1 \to m_1-1$ and $m_3 \to m_3-1$, we can identify three linear combinations that are $Q_\con$-closed: 
\begin{align}
& (\theta^{-+}_\min)_0 \ket{m_1, m_2, m_3} = \theta^{-+}_0 \ket{m_1, m_2, m_3} + \theta^{++}_0 \ket{m_1-1, m_2, m_3} ~,  \label{one-theta-qcon-exact-2} \\
& (\theta^{--}_\min)_0 \ket{m_1, m_2, m_3} = \theta^{--}_0 \ket{m_1, m_2, m_3} + \theta^{+-}_0 \ket{m_1-1, m_2, m_3} ~, \label{one-theta-qcon-exact-3}
\end{align}
and 
\be 
\ket{\theta} = \theta^{++}_0 \ket{m_1, m_2-1, m_3} + \theta^{+-}_0 \ket{m_1, m_2, m_3-1} \label{one-theta-qcon-exact-1} ~. 
\ee
The state \eqref{one-theta-qcon-exact-1} is $Q_\con$-exact since
\be 
\ket{\theta} = -Q_\con \, \theta_0^{++} \theta^{+-}_0 \ket{m_1, m_2, m_3}  ~, 
\ee
while the states \eqref{one-theta-qcon-exact-2} and \eqref{one-theta-qcon-exact-3} are $Q_\con$-exact because $\{ Q_\con, (\theta^{-+}_\min)_0 \} = \{ Q_\con, (\theta^{--}_\min)_0 \} = 0$ and we showed already that $\ket{m_1, m_2, m_3}$ is $Q_\con$-exact.

\paragraph{Two $\boldsymbol{\theta^{\alpha \beta}}$ modes:} We then consider the linear combination
\begin{multline}
    a_1 \, \theta^{++}_0 \theta^{--}_0 \ket{m_1-1, m_2, m_3} +
 a_2 \, \theta^{++}_0 \theta^{-+}_0 \ket{m_1-1, m_2 - 1, m_3 + 1} \\
 + a_3 \, \theta^{--}_0 \theta^{-+}_0 \ket{m_1, m_2, m_3} +
 a_4 \, \theta^{++}_0 \theta^{+-}_0 \ket{m_1 - 2, m_2, m_3} \\
 + a_5 \, \theta^{--}_0 \theta^{+-}_0 \ket{m_1-1, m_2 + 1, m_3-1} +
 a_6 \, \theta^{-+}_0 \theta^{+-}_0 \ket{m_1-1, m_2, m_3} ~, 
 \label{two-thetas-ansatz}
\end{multline}
where $a_1, \dots, a_6$ are arbitrary real coefficients. Requiring the state \eqref{two-thetas-ansatz} to be $Q_\con$-closed imposes 
\be 
\begin{aligned}
a_1 + a_3 + a_5 &=0 \,, &  a_2 + a_3 + a_6 &=0 ~,\\
a_4 + a_5 - a_6 &=0 \,, &  a_1 - a_2 - a_4 &=0 ~. 
\end{aligned}
\label{two-thetas-coeff}
\ee
We can thus consider the two linearly independent combinations, 
\begin{align}
    (\theta_\min^{--})_0 \, (\theta_\min^{-+})_0  \ket{m_1, m_2, m_3} & = - \theta^{++}_0 \theta^{--}_0 \ket{m_1-1, m_2, m_3} +
     \theta^{--}_0 \theta^{-+}_0 \ket{m_1, m_2, m_3}  \nn \\
    & \quad   - \theta^{++}_0 \theta^{+-}_0 \ket{m_1-2, m_2, m_3}
   - \theta^{-+}_0 \theta^{+-}_0 \ket{m_1-1, m_2, m_3} ~, 
    \label{thetamin-thetamin}
\end{align}
and 
\begin{multline}
\ket{\theta \theta} = \theta^{++}_0 \theta^{--}_0 \ket{m_1-1, m_2, m_3} +
\theta^{++}_0 \theta^{-+}_0 \ket{m_1-1, m_2 - 1, m_3 + 1} \\
- \theta^{--}_0 \theta^{+-}_0 \ket{m_1-1, m_2 + 1, m_3-1} 
- \, \theta^{-+}_0 \theta^{+-}_0 \ket{m_1-1, m_2, m_3} ~. 
\label{two-thetas-exact-state-2}
\end{multline}
The state \eqref{thetamin-thetamin} is $Q_\con$-exact because $Q_\con$ anti-commutes with $\theta_\min^{-\alpha}$ and $\ket{m_1, m_2, m_3}$ is itself $Q_\con$-exact, while \eqref{two-thetas-exact-state-2} is exact because
\begin{equation}
    \ket{\theta \theta} = Q_\con \left( \theta^{--}_0 \theta^{-+}_0 \theta^{+-}_0 \ket{m_1,m_2+1,m_3} + \theta^{++}_0 \theta^{--}_0 \theta^{-+}_0 \ket{m_1,m_2, m_3+1} \right) ~. 
\end{equation}

\paragraph{Three $\boldsymbol{\theta^{\alpha \beta}}$ modes:} Proceeding as in previous paragraphs, the only $Q_\con$-closed state we find is 
\begin{multline}
\ket{\theta \theta \theta} \equiv \theta^{--}_0 \theta^{-+}_0 \theta^{+-}_0 \ket{m_1, m_2, m_3 -1} 
+ \theta^{++}_0 \theta^{--}_0 \theta^{-+}_0 \ket{m_1, m_2-1, m_3} \\
- \theta^{++}_0 \theta^{-+}_0 \theta^{+-}_0 \ket{m_1-1, m_2-1, m_3} 
- \theta^{++}_0 \theta^{--}_0 \theta^{+-}_0 \ket{m_1-1, m_2, m_3-1} 
  ~. 
  \label{theta-theta-theta}
\end{multline}
Also this state is $Q_\con$-exact since
\be
\ket{\theta \theta \theta} = Q_\con \, \theta^{++}_0 \, \theta^{--}_0 \,  \theta^{+-}_0 \, \theta^{-+}_0 \ket{m_1, m_2, m_3}
~. 
\label{qcon-on-4-thetas}
\ee

\paragraph{Four $\boldsymbol{\theta^{\alpha \beta}}$ modes:}  None of the states
\be 
\theta^{++}_0 \, \theta^{--}_0 \, \theta^{-+}_0 \, \theta^{+-}_0 \ket{m_1, m_2, m_3}  
\ee
are $Q_\con$-closed. This simply follows from eq.~\eqref{qcon-on-4-thetas}. 

\subsection{Truncated representations}
\label{sec:truncated-reps}

In this section we consider the representations \eqref{zero-mode-rep} for $\lambda_2 \neq \tfrac12$ and $\lambda_3 = \tfrac12$, so that $m_3 \in \mathbb Z_{\leq 0}$ and 
\begin{equation}
    \pi^{--}_0 \ket{m_1, m_2, 0} = 0 ~. 
\end{equation}
We are going to show that also in this case, all the states that are closed are also exact. We proceed as in the previous section, analyzing states with increasing number of $\theta^{\alpha \beta}_0$ modes. 

\paragraph{No $\boldsymbol{\theta^{\alpha \beta}} $ modes:} For any $m_3 \in \mathbb Z_{\leq 0}$, eq.~\eqref{zero-thetas-Qcon-exact-2} still applies and hence $\ket{m_1, m_2, m_3}$ is exact. 

\paragraph{One $\boldsymbol{\theta^{\alpha \beta}} $ mode:} In addition to the states \eqref{one-theta-ansatz} with the choice of coefficients \eqref{one-theta-coeff} and $m_3 \in \mathbb Z_{<0}$ --- for which exactly the same analysis applies --- we also find the $Q_\con$-closed state
\be 
(\theta^{--}_\min)_0 \ket{m_1+1, m_2, 0} = \theta^{--}_0\ket{m_1, m_2,0} + \theta^{--}_0 \ket{m_1+1, m_2, 0} ~, 
\ee
which is exact because $\ket{m_1+1, m_2, 0}$ is exact and $\{Q_\con, (\theta^{--}_\min)_0 \} = 0$. 

\paragraph{Two $\boldsymbol{\theta^{\alpha \beta}}$ modes:} In addition to the states \eqref{two-thetas-ansatz} with the choice of coefficients \eqref{two-thetas-coeff} and $m_3 \in \mathbb Z_{<0}$, we find the $Q_\con$-closed state
\begin{align}
    (\theta_\min^{--})_0 \, (\theta_\min^{-+})_0  \ket{m_1, m_2, 0} & = - \theta^{++}_0 \theta^{--}_0 \ket{m_1-1, m_2, 0} +
     \theta^{--}_0 \theta^{-+}_0 \ket{m_1, m_2, 0}  \nn \\
    & \quad   - \theta^{++}_0 \theta^{+-}_0 \ket{m_1-2, m_2, 0}
   - \theta^{-+}_0 \theta^{+-}_0 \ket{m_1-1, m_2, 0} ~, 
   \label{thetamin-theamin-truncated}
\end{align}
which is exact because $\ket{m_1, m_2, 0}$ is itself exact. 

\paragraph{Three $\boldsymbol{\theta^{\alpha \beta}}$ modes:} In addition to the state \eqref{theta-theta-theta} with $m_3 \in \mathbb Z_{\leq 0}$, which is again exact, we do not find any $Q_\con$-closed state.

\paragraph{Four $\boldsymbol{\theta^{\alpha \beta}}$ modes:}  As one can directly check, none of the states
$\theta^{++}_0 \, \theta^{--}_0 \, \theta^{-+}_0 \, \theta^{+-}_0 \ket{m_1, m_2, m_3}$ with $m_3 \in \mathbb Z_{\leq 0}$ are $Q_\con$-closed.

\subsection{Doubly truncated representations}
\label{sec:doubly-truncated-reps}

Let us now consider the representations \eqref{zero-mode-rep} for $\lambda_2=\lambda_3=\tfrac12$, so that $m_2, m_3 \in \mathbb Z_{\leq 0}$ and 
\begin{equation}
    \pi^{-+}_0 \ket{m_1, 0, m_3} = 0 \,, \qquad \text{and} \qquad \pi^{--}_0 \ket{m_1, m_2, 0} = 0 ~. 
\end{equation}
We are going to show that the $Q_\con$ cohomology only contains the states listed in eq.~\eqref{non-min-reps}. Once more, we proceed considering an increasing number of $\theta^{\alpha \beta}$ zero modes.

\paragraph{No $\boldsymbol{\theta^{\alpha \beta}} $ modes:}
Unless $m_2 = m_3 = 0$, repeating the same analysis carried out in the previous sections, it is easy to show that the state $\ket{m_1, m_2, m_3}$ is exact. On the other hand, 
\be
\ket{m_1, 0, 0} 
\label{non-min-rep-0-singlet}
\ee
is closed but not $Q_\con$-exact. In fact, the $\lambda^{+\alpha}_0$ modes entering the definition of $Q_\con$, see eq.~\eqref{Qcon}, decrease the eigenvalues $m_2$ and $m_3$. On the other hand, since the representation truncates, $m_2 = m_3 =0$ is the maximal value that $m_2$ and $m_3$ can assume and therefore no state $\ket{\mathcal X}$ such that $Q_\con \ket{\mathcal X} = \ket{m_1, 0, 0} $ can be constructed for $\lambda_2=\lambda_3=\tfrac12$. Using the non-minimal free-field realization \eqref{keq1-nonmin-all}, one can verify that the states \eqref{non-min-rep-0-singlet} have $\mathfrak{sl}(2,\mathbb R)$ spin $j_\sl=0$ and are in the singlet representation of $\sutwo$. 

\paragraph{One $\boldsymbol{\theta^{\alpha \beta}}$ mode:} The states \eqref{one-theta-qcon-exact-2} and \eqref{one-theta-qcon-exact-3} are exact for any value of $m_2, m_3 \in \mathbb Z_{\leq 0}$ unless $m_2 = m_3 =0$. In this case, 
\begin{equation}
    (\theta^{-+}_\min)_0 \ket{m_1, 0, 0} \qquad \text{and} \qquad  (\theta^{--}_\min)_0 \ket{m_1, 0, 0}
    \label{the-thetas-not-exact}
\end{equation}
are still closed but no longer exact since $\ket{m_1, 0, 0}$ is not itself exact. One can check that the states \eqref{the-thetas-not-exact} have $\mathfrak{sl}(2,\mathbb R)$ spin $j_\sl=\frac{1}{2}$ and form the doublet representation of $\sutwo$. Finally, the state \eqref{one-theta-qcon-exact-1} is exact for any value of $m_3 \in \mathbb Z_{\leq 0}$. 

\paragraph{Two $\boldsymbol{\theta^{\alpha \beta}}$ modes:} The states \eqref{two-thetas-ansatz} with $m_2, m_3 \in \mathbb Z_{\leq -1}$ and obeying the relations \eqref{two-thetas-coeff} and the states \eqref{thetamin-theamin-truncated} with $m_2 \in \mathbb Z_{\leq -1}$ are exact by the same logic of the previous sections. There is only one more closed state one can construct,  which is
\begin{align}
    (\theta_\min^{--})_0 \, (\theta_\min^{-+})_0  \ket{m_1, 0, 0} & = - \theta^{++}_0 \theta^{--}_0 \ket{m_1-1, 0, 0} +
     \theta^{--}_0 \theta^{-+}_0 \ket{m_1, 0, 0}  \nn \\
    & \quad   - \theta^{++}_0 \theta^{+-}_0 \ket{m_1-2, 0, 0}
   - \theta^{-+}_0 \theta^{+-}_0 \ket{m_1-1, 0, 0} ~. 
    \label{thetamin-thetamin-doubly-truncated}
\end{align}
It is not exact because $\ket{m_1,0,0}$ is not exact. 

\paragraph{Three $\boldsymbol{\theta^{\alpha \beta}}$ modes:} All the closed states are given by eq.~\eqref{theta-theta-theta} with $m_2, m_3 \in \mathbb Z_{\leq 0}$. They are all exact. 

\paragraph{Four $\boldsymbol{\theta^{\alpha \beta}}$ modes:} 

None of the states $\theta^{++}_0 \, \theta^{--}_0 \, \theta^{-+}_0 \, \theta^{+-}_0 \ket{m_1, m_2, m_3}$ with $m_2, m_3 \in \mathbb Z_{\leq 0}$ are $Q_\con$-closed.

\section{The \texorpdfstring{$\mathfrak{psu}\boldsymbol{(1,1|2)_{n_5}}$}{} algebra}
\label{sec:algebras}

In this appendix we spell out our conventions for various commutation relations used in the main text. In our conventions, the $\mathfrak{psu}(1,1|2)_{n_5}$ algebra reads
\begin{subequations}
\begingroup
\allowdisplaybreaks
\begin{align}
    J^3(z) J^3(w) &= -\frac{n_5}{2(z-w)^2} ~, \\
    J^3(z) J^\pm(w) &= \pm \frac{J^\pm(w)}{z-w} ~, \\
    J^+(z) J^-(w) &= \frac{n_5}{(z-w)^2}  - \frac{2J^3(w)}{z-w} ~, \\
    K^3(z) K^3(w) &= \frac{n_5}{2(z-w)^2}  ~, \\
    K^3(z) K^\pm(w) &= \pm \frac{K^\pm(w)}{z-w}  ~, \\
    K^+(z) K^-(w) &= \frac{n_5}{(z-w)^2} + \frac{2 K^3(w)}{z-w} ~, \\
    J^a(z) S^{\alpha \beta  \gamma}(w) &= \frac{(\tau^a)^\alpha{}_\mu \, S^{\mu \beta \gamma}(w)}{2(z-w)} ~, \\
    K^a(z) S^{\alpha \beta  \gamma}(w) &= \frac{(\sigma^a)^\beta{}_\mu \, S^{\alpha \nu \gamma}(w)}{2(z-w)} ~, \\
    S^{\alpha \beta \gamma}(z) S^{\mu \nu \rho}(w) &=- \frac{n_5 \vareps^{\alpha \mu} \vareps^{\beta \nu} \vareps^{\gamma \rho}}{(z-w)^2} + \frac{\vareps^{\beta \nu} \vareps^{\gamma \rho} (\tau_a)^{\alpha \mu} J^a(w) - \vareps^{\alpha \mu} \vareps^{\gamma \rho} (\sigma_a)^{\beta \nu} K^a(w)}{z-w} ~, 
\end{align}
\endgroup
\label{psu112k}%
\end{subequations}
where Greek indices $\alpha, \beta, \dots$ take value in $\{+, -\}$, while the Lie algebra generators $\sigma^a$ and $\tau^a$ for $a \in \{+, -, 3 \}$ have been defined in eqs.~\eqref{sigma123} and \eqref{tau123}, respectively.

An alternative basis for $\mathfrak{psu}(1,1|2)_{n_5}$ is frequently adopted in the literature, including \cite{Berkovits:1999im}, and is related to the one introduced in \eqref{psu112k} as\footnote{Here we closely follow the conventions of \cite{Gaberdiel:2022als}, see also \cite{Gerigk:2012cq}.}
\begin{subequations}
\begin{align}
J^3 + K^3 &= - i K^{12} \,, & J^3 - K^3 &= - i K^{34} ~, \\
J^+ + J^- + K^+ - K^- &= 2 K^{23} \,, & J^+ + J^- - K^+ + K^- &= 2 K^{14} ~, \\
J^+ - J^- + K^+ + K^- &= 2 i K^{13} \,, & J^+ - J^- - K^+ - K^- &= -2i K^{24} ~,
\end{align}
and for $\alpha \in \{ +, - \}$
\begin{align}
S^{++\alpha} + S^{--\alpha} &= \sqrt 2 \, S^{1, \alpha} \,, & S^{++\alpha} - S^{--\alpha} &= -\sqrt 2 \, i \, S^{2,\alpha} ~, \\
S^{+-\alpha} + S^{-+\alpha} &= \sqrt 2\, i \, S^{3,\alpha} \,, & S^{+-\alpha} - S^{-+\alpha} &= \sqrt 2 \, S^{4,\alpha} ~. 
\end{align}
\label{alt-basis-psu}%
\end{subequations}
In terms of the generators \eqref{alt-basis-psu}, $\mathfrak{psu}(1,1|2)_{n_5}$ OPE's read 
\begin{align}
K^{ab}(z) K^{cd}(w) &= \frac{n_5 \, \vareps^{abcd}}{(z-w)^2}  + \frac{\delta^{ac} K^{bd}(w) - \delta^{ad}K^{bc}(w) - \delta^{bc}K^{ad}(w) + \delta^{bd}K^{ac}(w)}{z-w} ~, \\
S^{a,\alpha}(z) S^{b,\beta}(w) &= - \frac{n_5 \, \vareps^{\alpha \beta} \, \delta^{ab} }{(z-w)^2}  - \frac{\vareps^{\alpha \beta} \, \vareps^{abcd} K^{cd}(w)}{2(z-w)} ~, \\
K^{ab}(z) S^{c,\alpha}(w) &= \frac{\delta^{ac} S^{b,\alpha}(w) - \delta^{bc} S^{a,\alpha}(w)}{z-w} ~,  
\end{align}
where $K^{ab} = - K^{ba}$, the latin indices take values in $a,b,c,d \in \{1,2,3,4 \}$ and $\vareps^{+-} = 1$, $\vareps^{1234}=1$.

\section{Null gauging details}
\label{sec:null details}

\subsection{Modification of hybrid BRST}
\label{sec:null hyb brst}

In the map of null gauging to hybrid variables~\eqref{hybrid-map-12d}, the ghosts for the gauging of the fermionic null current map to the twisted ghosts $\gammattil,\betattil$.  Combined with the ghosts $\cnul,\bnul$, one has a twisted $\cN=2$ algebra with central charge~${\mathtt c}=-3$, whose generators are given by
\be
\begin{aligned}
\label{twisted null}
    \cT^{\rm gh}_n &= - \bnul\partial\cnul - \betattil\partial\gammattil ~, & \cJ^{\mathcal R}_n &= -\frac12 \betattil \gammattil ~,     \\
    \cG^{\rm gh, +}_n &= -\gammattil \bnul ~, & \cG^{\rm gh, -}_n &=  \betattil \partial \cnul ~. 
\end{aligned}
\ee

The BRST current including the contribution of the null gauging ghosts is given by 
\begin{equation}
    \widehat \mfJ_\brst = \widehat \mfJ_0 + \widehat \mfJ_1 + \widehat \mfJ_2 ~, 
\end{equation}
where 
\begin{align}
\widehat \mfJ_0 &= \big(\cT_{\mathfrak{psu}} \tight+ \cT_{\flat} \tight+ \cT^{\rm gh}_n \tight+ \cT_\rhob \big)\cstr \tight+ \partial \cstr (\bstr \cstr) - \cG_{\flat}^+ \!- \cG^{\rm gh, +}_n 
\!-\partial \Big(\partial \rhob \, \cstr -\partial \cstr + 2 \big(\cJ^{\mathcal R}_{\flat} + \cJ^{\mathcal R}_{n}\big) \, \cstr  \Big) ~, 
\nn \\[.2cm]
\widehat \mfJ_1 &= \mfJ_1 =  \frac{e^{-\rhob}}{2 \, \sqrt{\nfive}}\Big[ \big(\vareps_{\mu\nu} \, (\tau_a)_{\alpha\beta} \, J^a +  \vareps_{\alpha\beta} \, (\sigma_a)_{\mu\nu} \, K^a \big)p^{\alpha\mu}p^{\beta\nu} - 4 \, \vareps_{\alpha \beta} \, \vareps_{\mu \nu} \, p^{\alpha\mu}\partial p^{\beta\nu} \Big]    ~, 
 \label{hyb-BRST-before-similarity-null}
\\[.2cm]
\widehat \mfJ_2 &= -e^{-2\rhob} p^{++} p^{+-} p^{-+} p^{--} \big( \cG_{\flat}^- + \cG^{\rm gh, -}_n + \bstr \big)  ~.  \nn
\end{align} 
Here, the twisted flat directions have been extended to include an extra $\bR^{1,1}$ parametrized by $X^{~}_0,X^*_0,\Psi^{~}_0,\Psi^*_0$ in addition to $\bT^4$ (which is now allowed to be decompactified to $\bR^4$).

Let us now perform a similarity transformation~\eqref{simtransf} where the current $R$ now gets promoted~to 
\begin{equation}
    \label{null similarity}
    \widehat{R} = \oint\! dz \, \cstr \big(  \cG_{\flat}^- + \cG^{\rm gh, -}_n \big) \,. 
\end{equation}
After the similarity transformation, the $\mathcal N=2$ algebra reads 
\begin{align}
    \widehat{\cT} &= \cT_{\mathfrak{psu}} + \mathcal T_{\flat} + \mathcal T_\rhob  +  T_{\bstr \cstr} + \cT^{\rm gh}_n ~, \\
    \widehat{\cG}^+ &= \widehat \mfJ_0 + \widehat \mfJ_1 + \widehat \mfJ_2 ~, \\
    \widehat{\cG}^- &= \bstr - \cG_{\flat}^- - \cG^{\rm gh, -}_n ~, \\
    \widehat{\cJ}^{\, \mathcal R} &= \frac{1}{2} (\partial \rhob + i \partial \sigmab) +  \cJ^{\mathcal R}_{\flat} +  \cJ^{\mathcal R}_{n} ~, 
\end{align}
where 
\begin{subequations}
\label{Jhat hybrid}
\begin{align}
\widehat \mfJ_0 &= \Big[\cT_{\mathfrak{psu}} - \hf \partial(\rhob\tight+ i\sigmab) \partial(\rhob\tight+ i\sigmab) + \hf\partial^2(\rhob\tight+ i \sigmab) \Big] e^{i\sigmab} - \cG_{\flat}^+ - \cG^{\rm gh, +}_n ~, 
\label{J0hat hybrid}\\[.2cm]
\widehat \mfJ_1 &= \mfJ_1 = \frac{e^{-\rhob}}{2 \, \sqrt{\nfive}}\Big[ \big(\vareps_{\mu\nu} \, (\tau_a)_{\alpha\beta} \, J^a +  \vareps_{\alpha\beta} \, (\sigma_a)_{\mu\nu} \, K^a \big)p^{\alpha\mu}p^{\beta\nu} - 4 \, \vareps_{\alpha \beta} \, \vareps_{\mu \nu} \, p^{\alpha\mu}\partial p^{\beta\nu} \Big]  ~, 
\label{J1hat hybrid}
\\[.2cm]
\widehat \mfJ_2 &= \mfJ_2 = - e^{-2\rhob-i \sigmab} p^{++} p^{+-} p^{-+} p^{--}   ~.
\label{J2hat hybrid}
\end{align}
\end{subequations}
One naturally extends the topologically twisted $\bT^4$ contribution to include the contributions of $\bR^{1,1}$ and the twisted ghosts $\betattil,\gammattil,\bnul,\cnul$, which don't change the algebra~\eqref{N=2-twisted-T+} (which still has central charge ${\mathtt c}=6$).

The RNS null gauging BRST charge $Q_\nul$ undergoes a sequence of transformations, first the mapping~\eqref{psi3+chi3} to hybrid variables, followed by the similarity transformation~\eqref{simtransf} generated by $\widehat{R}$ above.  The end result is
\begin{equation}
    Q_\nul ~\longrightarrow~ \widehat Q_\nul = Q_\nul + \sqrt{n_5}\, \oint\! dz \, \big( \partial \cnul \cstr e^{\rhob} \theta^{++} \theta^{--}   \big)(z) ~. 
\end{equation}

\subsection{Picture changing}
\label{sec:null pics}

Spin fields in the RNS formalism implement $\bZ_2$ twisted holonomy for all worldsheet spinors in the matter sector $\psi_a,\chi_a,\zeta_\sfj, \zeta_\sfj^*$, as well as the ghosts $\gammastr,\betastr,\gammanul,\betanul$.  They are most conveniently written in bosonized form, as in eqs.~\eqref{hybrid map}, \eqref{hybrid-map-12d}.  

Each physical vertex operator has a variety of equivalent representatives in different charge sectors of the bosonized spinor ghosts.  In the presence of null gauging, there are pictures for $\phinul$ as well as the usual $\phistr$.  While the simplest, canonical picture in which to write Ramond vertex operators for the string ghosts is the $(-1/2)$ picture, for the null gauging ghosts it is the $(+1/2)$ picture~\rcite{Martinec:2022okx}.  In the conventions of this paper, one finds that while the spin fields $p^{++},p^{--}$ of eq.~\eqref{hybrid-map-12d} commute with the bosonic null constraint $\oint\!\cnul\cJ$, they are not invariant under the fermionic null constraint $\oint\!\gammanul\Lambda$ (this is obvious in the hybrid representation~\eqref{psi3+chi3}). 

As with the usual string pictures under $\phistr$, there is a picture changing operation for $\phinul$.  If we label vertex operators by their $(\phistr,\phinul)$ pictures, then
\be
\cO^{(q_s+1,q_n)}(z) = \big\{ Q_\str,\xistr(z)\cO^{(q_s,q_n)}(z) \big\}
~,\qquad
\cO^{(q_s,q_n+1)} = \big\{ Q_\nul,\xinul(z)\cO^{(q_s,q_n)}(z) \big\}
\ee

One can then find the operators in the $(-1/2,-1/2)$ picture equivalent to $p^{+-}$ and $p^{-+}$ defined in the canonical $(-1/2,+1/2)$ picture, \eg\ for $p^{+-}$
\begin{align}
\check p^{+-} \equiv \big(p^{+-}\big)^{(-\half,-\half)} = e^{\half[(+H_1-H_2)+(H_4+H_5+H_6)-\phistr]}\, \Big( e^{\half(-H_3-\phinul)}+\cnul\partial\xinul\,e^{\half(+H_3-3\phinul)} \Big)  ~,
\end{align}
and similarly for $p^{-+}$.  Here the second term is annihilated by the picture changing operation, but is required for BRST invariance.

Considering the two simplest Ramond pictures $\pm\half$ for the null ghost $\phinul$, one has the spin fields
\be
\begin{aligned}
\cO_1 &= \exp\big[\hf\big( \vareps_1(H_1-H_2)+H_3+\vareps_{456}\cdot H_{456}-\phistr+\phinul \big)\big] ~, \\[.1cm]
\cO_2 &= \exp\big[\hf\big( \vareps_1(H_1-H_2)+H_3+\vareps^*_{456}\cdot H_{456}-\phistr-\phinul \big)\big] ~, \\[.1cm]
\cO_3 &= \exp\big[\hf\big( \vareps_1(H_1-H_2)-H_3+\vareps^*_{456}\cdot H_{456}-\phistr+\phinul \big)\big]+\dots ~,  \\[.1cm]
\cO_4 &= \exp\big[\hf\big( \vareps_1(H_1-H_2)-H_3+\vareps_{456}\cdot H_{456}-\phistr-\phinul \big)\big]+\dots ~,
\end{aligned}
\ee
where in the last two expressions, the $\dots$ are additional terms needed for invariance under the $\gammastr (G_{\mathrm m} + \tfrac12 G^\gh_s + G^\gh_n) $ term in $Q_\str$; also, $\vareps_{456}$ and $\vareps^*_{456}$ are Dynkin labels for 6d spinors of opposite chirality.

In the analysis of BRST cohomology, one finds that $\cO_1$ is in the BRST cohomology, while $\cO_4$ is related to $\cO_1$ by null-ghost picture changing.  On the other hand, $\cO_3$ is not invariant under $Q_\nul$, while $\cO_2$ is $Q_\nul$-exact.  Thus we have chosen operators of the form $\cO_1$ in the analysis of null gauging in Sections~\ref{sec:nullgauge} and~\ref{sec:supertube}.


\bibliographystyle{JHEP}      

\bibliography{fivebranes}


\end{document}
